\documentclass[letterpaper]{article} 
\usepackage{aaai25}  
\usepackage{times}  
\usepackage{helvet}  
\usepackage{courier}  
\usepackage[hyphens]{url}  
\usepackage{graphicx} 
\usepackage{natbib}  
\usepackage{caption} 
\usepackage{algorithm}
\usepackage{algorithmic}

\usepackage{newfloat}
\usepackage{listings}
\DeclareCaptionStyle{ruled}{labelfont=normalfont,labelsep=colon,strut=off} 
\floatstyle{ruled}
\newfloat{listing}{tb}{lst}{}
\floatname{listing}{Listing}
\usepackage[dvipsnames]{xcolor}
\usepackage{amsmath} 
\usepackage{subfigure}
\usepackage{multirow}

\title{Disrupt Your Research Using Generative AI Powered ScienceSage}
\author {
    Yong Zhang\textsuperscript{\rm 1},
    Eric Herrison Gyamfi\textsuperscript{\rm 2},
    Kelly Anderson\textsuperscript{\rm 1},
    Sasha Roberts\textsuperscript{\rm 1},
    Matt Barker\textsuperscript{\rm 1}
}
\affiliations {
    \textsuperscript{\rm 1} Corporate Functions R\&D, Discovery\& Innovation Platforms, P\&G \thanks{This paper has been accepted by Workshop of Deployable AI at AAAI 2025.}\\
    \textsuperscript{\rm 2} Department of Mathematical Sciences, Division of Statistics and Data Science, University of Cincinnati \\
    
    \{zhang.y.13, anderson.kl.1, roberts.sa, barker.ml\}@pg.com, gyamfien@mail.uc.edu 
}

\usepackage{bibentry}

\begin{document}

\maketitle

\begin{abstract}

Large Language Models (LLM) are disrupting science and research in different subjects and industries. Here we report a minimum-viable-product (MVP) web application called \textbf{ScienceSage}. It leverages generative artificial intelligence (GenAI) to help researchers disrupt the speed, magnitude and scope of product innovation. \textbf{ScienceSage} enables researchers to build, store, update and query a knowledge base (KB). A KB codifies user’s knowledge/information of a given domain in both vector index and knowledge graph (KG) index for efficient information retrieval and query. The knowledge/information can be extracted from user’s textual documents, images, videos, audios and/or the research reports generated based on a research question and the latest relevant information on internet. The same set of KBs interconnect three functions on \textbf{ScienceSage}: ‘Generate Research Report’, ‘Chat With Your Documents’ and ‘Chat With Anything’. We share our learning to encourage discussion and improvement of GenAI's role in scientific research.
\end{abstract}

%
\begin{links}
    \link{ScienceSage Github}{https://github.com/zhy5186612/GenAI-ResearchAssistant-ScienceSage}
\end{links}

\section{Introduction}

Large Language Models (LLM) are disrupting science and research in different subjects and industries \citep{wang2023scientific}. As an example, ChemCrow uses GPT-4 to integrate 18 expert-designed tools to accomplish tasks across organic synthesis, drug discovery and materials design \citep{m2024augmenting}. There are also tools like ‘REINVENT 4’ specifically designed for a narrow scientific research task. ‘REINVENT 4’ leverages generative AI framework to design small molecules for drug discovery \citep{loeffler2024reinvent}.      

LLMs have great potential to disrupt product innovation, consumer research, marketing, and business models etc in consumer packaged goods (CPG) and other industries. These innovation activities generally involve searching, retrieving and summarizing domain specific information and knowledge embedded in multimodal data including text, image, audio and video. The multimodal data could be publicly available on internet and/or proprietary to a specific institution. Retrieval-Augmented Generation (RAG) \citep{guu2020realmretrievalaugmentedlanguagemodel} is commonly used to retrieve relevant information from user’s documents and then generate an response to answer a query \citep{chen2023benchmarkinglargelanguagemodels}.

A regular RAG can indeed generate a response to answer user's question. However, most of time the answer is a simple paragraph. A researcher may want to have more details about his query. It could be a reference to verify the answer, or facts and mechanism to interpret the answer. Rather than a simple answer, a researcher may also want to have a structural report containing title, sub-title, discussion, conclusion and references etc.       

Furthermore, the information and knowledge contained in LLMs or user's data can be outdated. Knowledge from LLMs like ChatGPT generally has a cutoff time for obtaining the training data. Likewise for user's multimodal data, as an example, a researcher may not be aware of some of the latest articles or technical reports on a given topic. We need capabilities and tools to quickly retrieve the latest relevant information and digest this information for insights.

We designed and built a MVP web application \textbf{ScienceSage} to address the above challenges. We first elaborate on the architecture and functionality of this MVP web application. We then proceed to show our evaluations on the RAG component. After briefly mentioning the usage of \textbf{ScienceSage}, we finally discuss challenges, future improvement plans and conclude the article. 

\section{ScienceSage web application}
\subsection{User interface}
\begin{figure}[h!] 
    \centering
        \includegraphics[scale=0.25]{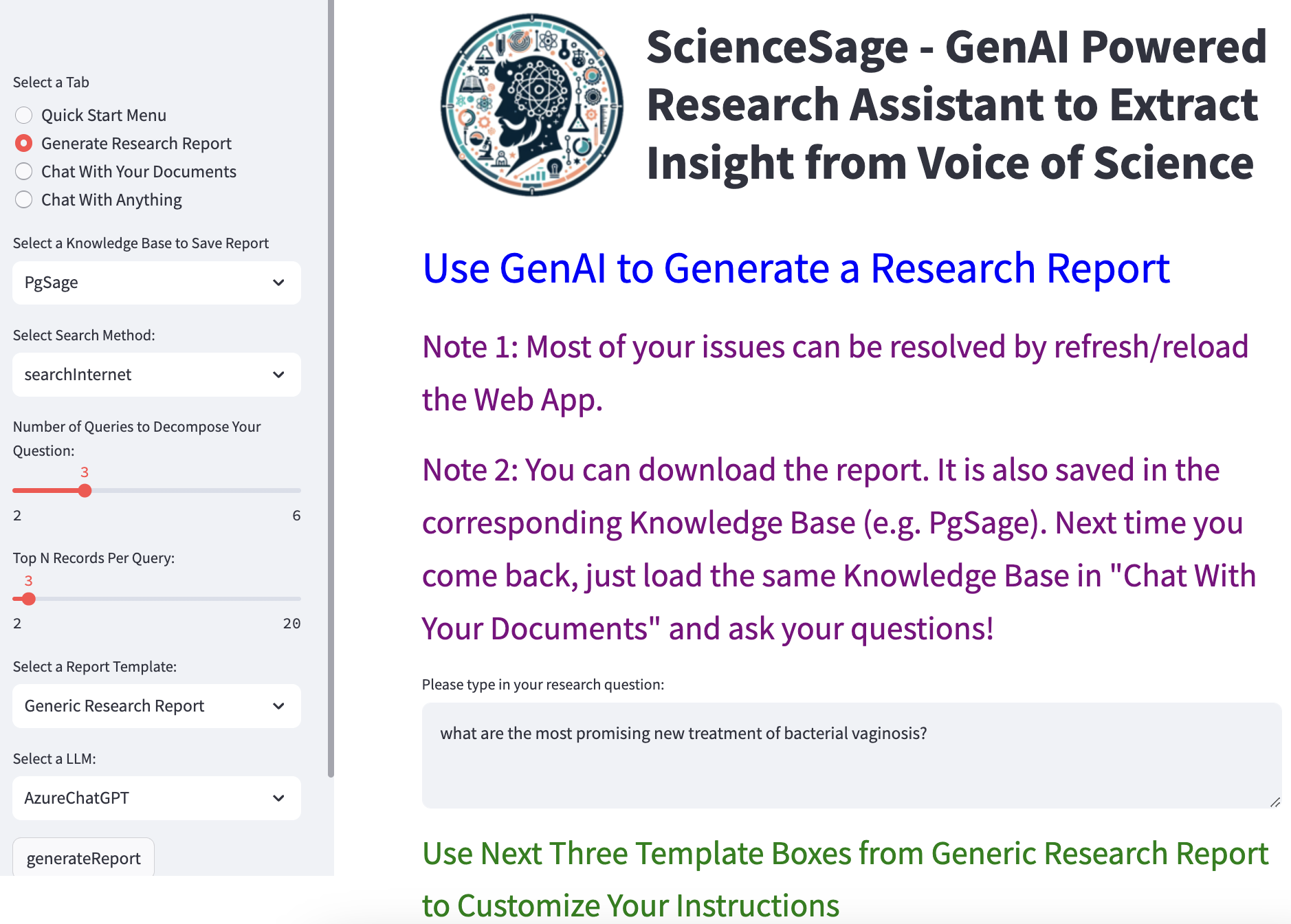}
    \caption{\textbf{ScienceSage} web application}
    \label{fig:webApp}
\end{figure}

Figure \ref{fig:webApp} shows the user interface of \textbf{ScienceSage}. The ‘Quick Start Menu’ provides high level explanation of \textbf{ScienceSage} and example questions, demo data, and video to enable quick test use and DIY applications. The 'Generate Research Report' tab can generate a comprehensive research report based on a user's research question and the latest information on internet. \textbf{ScienceSage} enables users to edit the built-in prompt template to quickly customize their own report without worrying about the coding and information retrieval. 

The 'Chat With Your Documents' tab enables a user to upload their own documents and then chat with a RAG to get quick and concise answers with references and/or a knowledge graph from these documents. The ‘Chat With Anything’ enables a user to upload multimodal data (text, image, audio \& video) and then chat with them to get insights. A common set of knowledge bases are used to interconnect ‘Generate Research Report’, ‘Chat With Your Document’ and ‘Chat With Anything’. Users can choose to use an existing KB or to create a new KB as needed. Users can chat with that KB, also store and update knowledge in that KB. The knowledge can be extracted from either user’s multimodal data or research reports generated by \textbf{ScienceSage}. Users don't have to upload all multimodal data all at once. Practically, it is good to build and update the KB incrementally as it both smooth computational burden and ease user's effort to curate all data at once. 

For generating a research report, \textbf{ScienceSage} always searches the internet or scientific databases to get the latest information.  We designed multi-stage prompts to instruct the LLM to generate a structural and comprehensive research report. It has a hierarchical structure such as title, subtitles, conclusion etc. It also lists the top relevant references. \textbf{ScienceSage} first decomposes the question into a number of queries and then searches the internet. It then return the top relevant urls where the contents are scraped. The scraped websites are then summarized. Finally a comprehensive and structural report is generated using the summaries with the urls as references. Users can set the number of queries and the number of top relevant websites. Users can also limit the search to scientific papers in arXiv database. In that case, the relevant papers are retrieved, summarized, referenced and used to generate the report. The generated report is downloadable and is also automatically saved and indexed in the corresponding KB.    

The 'Chat With Your Documents' is achieved through a RAG which can only consume textual data. Users can choose to use a RAG based on vector index, knowledge graph (KG) index or a custom index combining both vector index and KG index. These indices are saved in the corresponding KB selected by users. All three indices are detailed in the section of RAG evaluation. 

The ‘Chat With Anything’ is achieved through a multimodal RAG. It builds a multimodal index based on any single modality or combination of modalities of data. For now, this multimodal index is session based and we are testing to store and update it in LanceDB. \textbf{ScienceSage} does not save the picture, audio and video data. Instead, it saves indexes of the transcribed text data from both audio and video data in the corresponding KB selected by users. 

\subsection{Architecture}
\begin{figure}[h!] 
    \centering
        \includegraphics[scale=0.41]{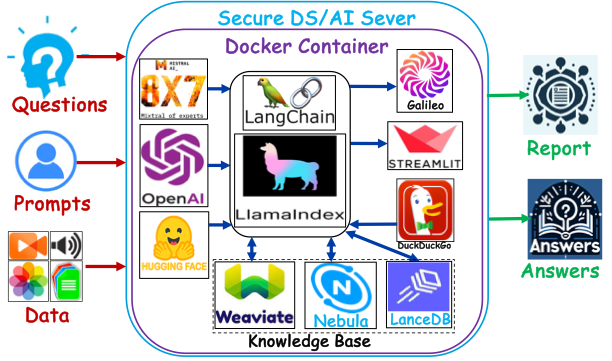}
    \caption{Architecture of \textbf{ScienceSage}}
    \label{fig:RagA}
\end{figure}

Figure \ref{fig:RagA} depicts the architecture of \textbf{ScienceSage}. Streamlit was used to develop the front end user interface in figure \ref{fig:webApp}. \textbf{ScienceSage} currently supports ChatGPT4 and GPT4-Vision through Azure and a privately deployed LLM Mixtral\_8X7B. These are some of the best performing commercial and open source LLMs, respectively. It uses the multilingual Hugging Face all-distilroberta-v1 model as its text embedding model \citep{wolf-etal-2020-transformers}, and open source CLIP model as its image embedding model \citep{wolf-etal-2020-transformers}. 

A Langchain \citep{Chase_LangChain_2022} agent was implemented to perform the tasks of 'Generate Research Report'. LlamaIndex \citep{Liu_LlamaIndex_2022} RAGs were implemented to perform tasks of 'Chat With Your Documents' and ‘Chat With Anything’. The extracted knowledge is stored as indices in Weaviate vector database and Nebula graph database . \textbf{ScienceSage} uses DuckDuckGoSearchAPI and ArxivRetriever to search internet and Arxiv databases, respectively. We plan to add more scientific databases in the back end in future. We also use Galileo to monitor the usage and improve different functionalities based on usage data. We use Weaviate and Nebula Graph to store the vector embedding and graph embedding, respectively. LanceDB is under test to store multimodal embedding and index.

Mixtral\_8X7B is an open source LLM and free to use. We deployed the Mixtral\_8X7B as a private LLM on a GPU server on premise. We have total control on this private LLM and the generated contents. These generated contents and user's documents never reach to any external parties. Additionally, Microsoft filters out and block your answers/response if it thinks your input or output is offensive. As a non-consumer facing and internal research tool, we don't have filtering policy on Mixtral\_8X7B. Users are guaranteed to have a valid response. \textbf{ScienceSage} currently supports multiple files in multiple formats including .pdf, .docx, .txt, .csv and .xlsx etc.

\section{RAG evaluation}

\subsection{RAG based on vector index}
Embedding is numerical representations of entities (like words or objects) in a continuous, high-dimensional space \citep{Zamani2016EstimatingEV}. Embedding is the digital fingerprint for words or concepts. It captures the context and semantics of these entities based on their occurrence in large text corpora. Vector index can be built based on embeddings. 

\subsection{RAG based on knowledge graph index}
A knowledge graph is a structured representation of knowledge in the form of entities and their relationships, typically organized as triples (Subject, Predicate, Object). This structure allows for explicit, logical representation and reasoning about the relationships between entities \citep{7498307}. These triples define the relationships between entities. LlamaIndex and Nebula graph were used to create and manage these graphs from textual data. This process also creates KG index for a RAG. 

\subsection{RAG based on both vector and KG index}
Custom index is a hybrid structure that combines the strengths of vector embedding and knowledge graphs. It leverages the entities in a knowledge graph and the relevant vector embeddings, enabling both fast similarity searches and deep logical reasoning \citep{mittal2017thinkingfastslowcombining}. 

To answer a user query, this custom RAG first retrieves information from vector index and KG index, separately. It then combines all retrieved information together and finally synthesizes an a response \citep{Liu_LlamaIndex_2022}.   

\subsection{Numerical experiment}

A detailed numerical study was conducted to evaluate the effectiveness of RAGs based on three different indexing methods: vector index, KG index, and a custom index mentioned early. The study assesses the accuracy and quality of responses generated by these methods across queries of differing complexity: easy, medium, and hard.

\subsection{Dataset}

The dataset \cite{DataSource2008} in our study is meticulously designed to explore the performance of different indexing methods across a range of query difficulties and keyword occurrences. The queries are divided into three levels of difficulty: easy, medium, and hard, based on the complexity and specificity of the information requested. Each query, regardless of its difficulty level, is processed through three indexing methods. 

Three different difficulty levels of queries are described below.
\begin{itemize}
    \item \textbf{Easy queries}: These queries are straightforward with clear intent and well-defined keywords that directly match relevant information in the dataset. Easy query serves as a benchmark to test the basic retrieval capabilities of each indexing method.
    \item \textbf{Medium queries}: These queries involve moderate complexity, potentially requiring some level of inference, ambiguity resolution, or the retrieval of information from multiple related entities or concepts. Medium queries assess the ability of each indexing method to handle more challenging information retrieval tasks.
    \item \textbf{Hard queries}: These queries are the most complex, often involving vague, abstract, or highly specific information that may not be directly referenced in the dataset. Hard queries test the robustness and depth of understanding that each indexing method can provide.
\end{itemize}

This study also categorizes queries based on the keyword occurrences, which substantially influences the performance of each indexing method.  
\begin{itemize}
    \item \textbf{High occurrence}: Keywords are abundantly present in the query, making it easier for the indexing methods to match the query with relevant data. This scenario evaluates how effectively each method leverages explicit cues in the queries.
    \item \textbf{Medium occurrence}: Keywords are present but not abundant. This reflects a typical real-world query where key terms are included, but the query might require some interpretation beyond the explicit keywords.
    \item \textbf{Low occurrence}: Keywords are sparse or minimally represented in the query, challenging the indexing methods to infer meaning and to retrieve relevant information based on broader context rather than direct keyword matching.
\end{itemize}

Table \ref{datades} lists the size and example queries at each difficulty levels and keyword occurrences.   

\begin{table*}[h!]
\centering
\begin{tabular}{|c|c|c|c|p{6cm}|}
\hline
\textbf{Query Size} & \textbf{Keyword Occurrence} & \textbf{Query Difficulty} & \textbf{Query Size} & \textbf{Sample of queries} \\ \hline
\multirow{3}{*}{765} & \multirow{3}{*}{Low occurrence} & Easy & 255 & In which year was the 16th President of the USA born? \\ \cline{3-5} 
 &  & Medium & 255 & What were the main aspects of the Proclamation issued to free slaves by the 16th President? \\ \cline{3-5} 
 &  & Hard & 255 & Evaluate the influence of the 16th President on Union military tactics during the Civil War. \\ \hline
\multirow{3}{*}{765} & \multirow{3}{*}{Medium occurrence} & Easy & 255 & What year was Abraham Lincoln born?\\ \cline{3-5} 
 &  & Medium & 255 & What were the key points of the Emancipation Proclamation issued by Abraham Lincoln?\\ \cline{3-5} 
 &  & Hard & 255 & Analyze Abraham Lincoln's impact on the Civil War strategies of the Union.\\ \hline
\multirow{3}{*}{765} & \multirow{3}{*}{High occurrence} & Easy & 255 & "In which exact year was Abraham Lincoln, the 16th President of the United States who led the country through the Civil War and issued the Emancipation Proclamation, born, considering his early life in Kentucky and his rise in Illinois politics? \\ \cline{3-5} 
 &  & Medium & 255 &  What were the key points of the Emancipation Proclamation, issued by Abraham Lincoln during the American Civil War, and how did it influence the abolition of slavery in the Confederate states, the reaction of the Union army, and the broader implications for civil rights in the United States?\\ \cline{3-5} 
 &  & Hard & 255 & Analyze Abraham Lincoln's impact on the Civil War strategies of the Union, including his decisions as Commander-in-Chief, his interactions with generals like Ulysses S. Grant and William Tecumseh Sherman, his role in the implementation of the Anaconda Plan, and his influence on military tactics that led to key victories such as the Battle of Gettysburg and the Siege of Vicksburg.  \\ \hline
\end{tabular}
\caption{Description of queries for evaluating the quality of RAG response based on 85 different documents.}
\label{datades}
\end{table*}

\subsection{Evaluation metrics}
LlamaIndex offers LLM-based evaluation modules to measure the quality of results. This uses a "gold" LLM (e.g. GPT-4) to decide whether the predicted answer is correct in a variety of ways. We used correctness, relevance and faithfulness / hallucination to evaluate different RAGs. There metrics are defined as below. 

\begin{enumerate}
\item Correctness: whether the generated answer matches that of the reference answer given the query (requires labels).This metric uses the CorrectnessEvaluator in LlamaIndex \citep{touvron2023llama2openfoundation} to evaluate the correctness of a generated answer against a reference answer. The score has to be between 1 and 5, where 1 is the worst and 5 is the best: if the generated answer is not relevant to the user query, you should give a score of 1; if the generated answer is relevant but contains mistakes,you should give a score between 2 and 3;if the generated answer is relevant and fully correct, you should give a score between 4 and 5.

\item Faithfulness or hallucination: faithfulness is a measure of whether the generated answer is faithful to the retrieved contexts. In other words, it measures whether there is any hallucination in the generated answer. This uses the FaithfulnessEvaluator to measure if the response from a query engine matches any source nodes. This is useful for measuring if the response was hallucinated. The evaluator returns a score between 0 and 1, where 1 means the response is faithful to the retrieved contexts. Given a generated answer \( \text{as}(q) \) for a question \( q \), the faithfulness metric \( F \) is computed as follows: Decompose the answer \( \text{as}(q) \) into a set of individual statements \( S(\text{as}(q)) \), where \( S(\text{as}(q)) = \{s_1, s_2, \ldots, s_n\} \).For each statement \( s_i \) in the set \( S(\text{as}(q)) \), determine whether it can be inferred from the retrieved context \( c(q) \) using the verification function \( v(s_i, c(q)) \).  The faithfulness score \( F \) is formulated as in equation \ref{eq: faithfulness} :
\begin{equation}
F = \frac{|V|}{|S|}
\label{eq: faithfulness}
\end{equation} 

where \( |V| \) is the number of statements \( s_i \) in \( S(\text{as}(q)) \) that are verified as supported by the context \( c(q) \) and \( |S| \) is the total number of statements in \( S(\text{as}(q)) \).

\item Relevancy: whether retrieved context is relevant to the query. The metric uses RelevancyEvaluator to measure if the response and  source nodes match the query. This is useful for measuring if the query was actually answered by the response. The evaluator returns a score between 0 and 1, where 1 means the response is relevant to the query otherwise not relevant. The metric is giving by 
\begin{equation}
Relevance = \frac{\text{\# Relevant Concepts in Response}}{\text{Total \# Concepts in Response}} 
\end{equation}

\end{enumerate}

Mean and standard deviation were calculated for each evaluation metric at each difficulty level and each keyword occurrence using the corresponding queries. Higher bars mean better and shorter error bar indicates that the responses metrics are consistently close to the mean, suggesting reliability. A higher error bar suggests greater variability and potential inconsistency in the retrieval process.

\section{Results}

\subsection{Correctness} 
As shown in figure \ref{fig:evalDLa}, across easy, medium, and hard queries, the RAG based on custom index consistently achieves the highest correctness scores. This can be attributed to the integration of two complementary strengths: the vector index’s ability to capture broad semantic context and the KG index’s ability to anchor responses in specific entities and factual information. For easy queries, where the intent is clear and the required information is straightforward, the vector index performs reasonably well, leveraging its semantic understanding. However, as query difficulty increases (medium and hard queries), the vector index’s correctness declines because the complexity of the information needed outstrips its capacity to ensure factual accuracy without rich and diverse information retrieval. The KG index, with its focus on keywords/entities, shows better correctness for medium and hard queries compared to the vector index, as it can more reliably retrieve factual data associated with specific entities. However, it still falls short particularly in complex scenarios where both context and factual precision are crucial.

After aggregating across difficulty levels and keywords occurrences, the RAG based on custom index again emerges as the superior method (figure A1(a)). This consistent performance across different query sets demonstrates the robustness of the combined approach in handling a wide variety of query types. The Vector Index, while showing improvement with simpler queries, continues to struggle with maintaining correctness in harder queries due to its reliance on semantic similarity, which is less effective when complex or nuanced information is required. The Knowledge Index shows relatively stable correctness but cannot match the Combined Index’s performance, especially in scenarios requiring both contextual understanding and precise factual retrieval.

\subsection{Relevance}
\begin{figure}[!ht] 
    \centering
    \subfigure[Correctness of response]{
        \includegraphics[width=0.9\linewidth]{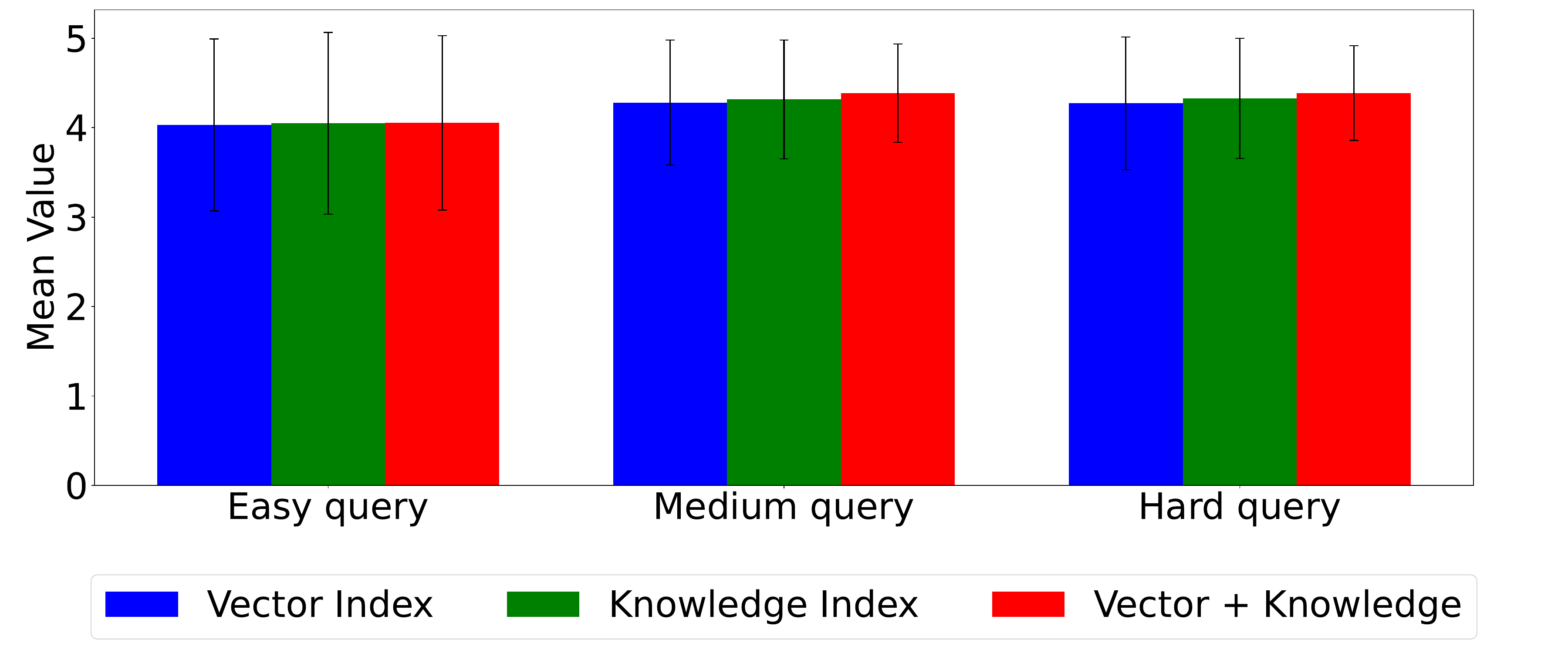}
        \label{fig:evalDLa}}
    \subfigure[Relevance of response]{
        \includegraphics[width=0.9\linewidth]{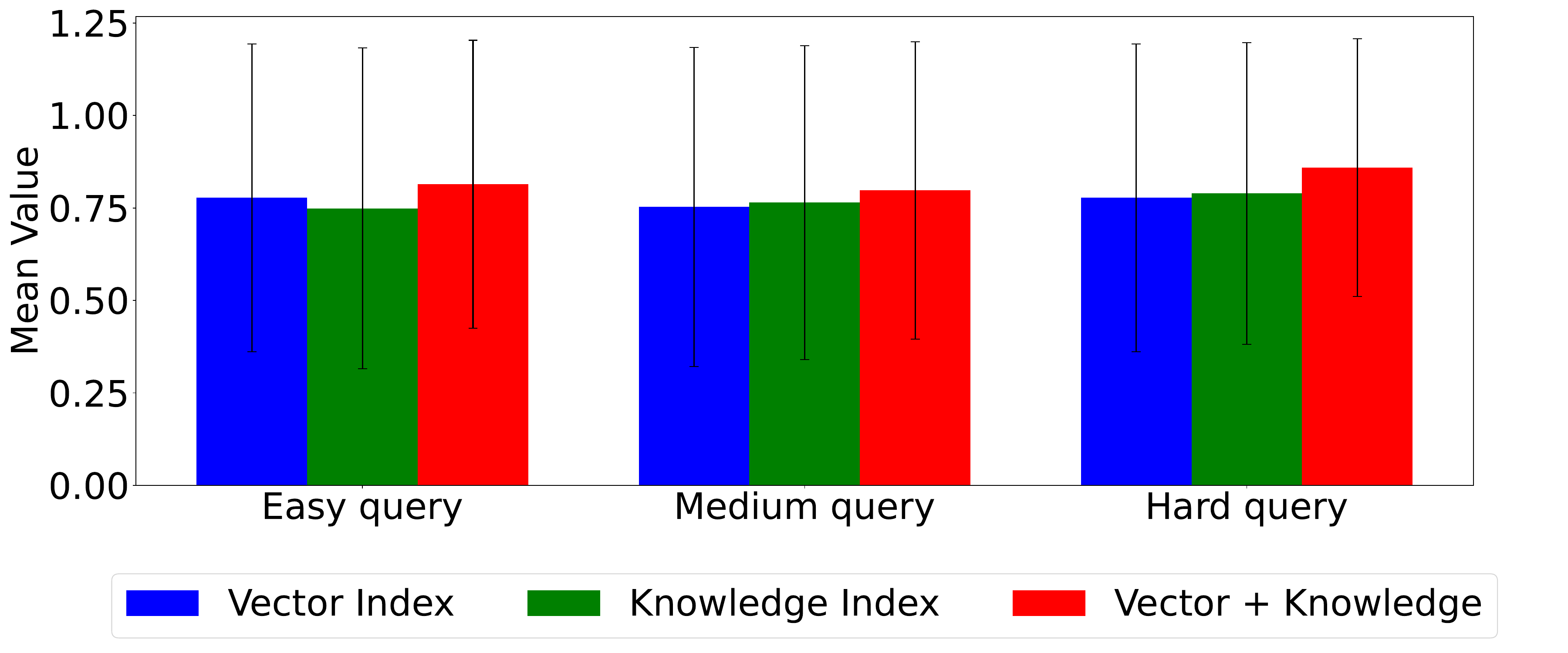}
        \label{fig:evalDLb}}
    \subfigure[Faithfulness of response]{
        \includegraphics[width=0.9\linewidth]{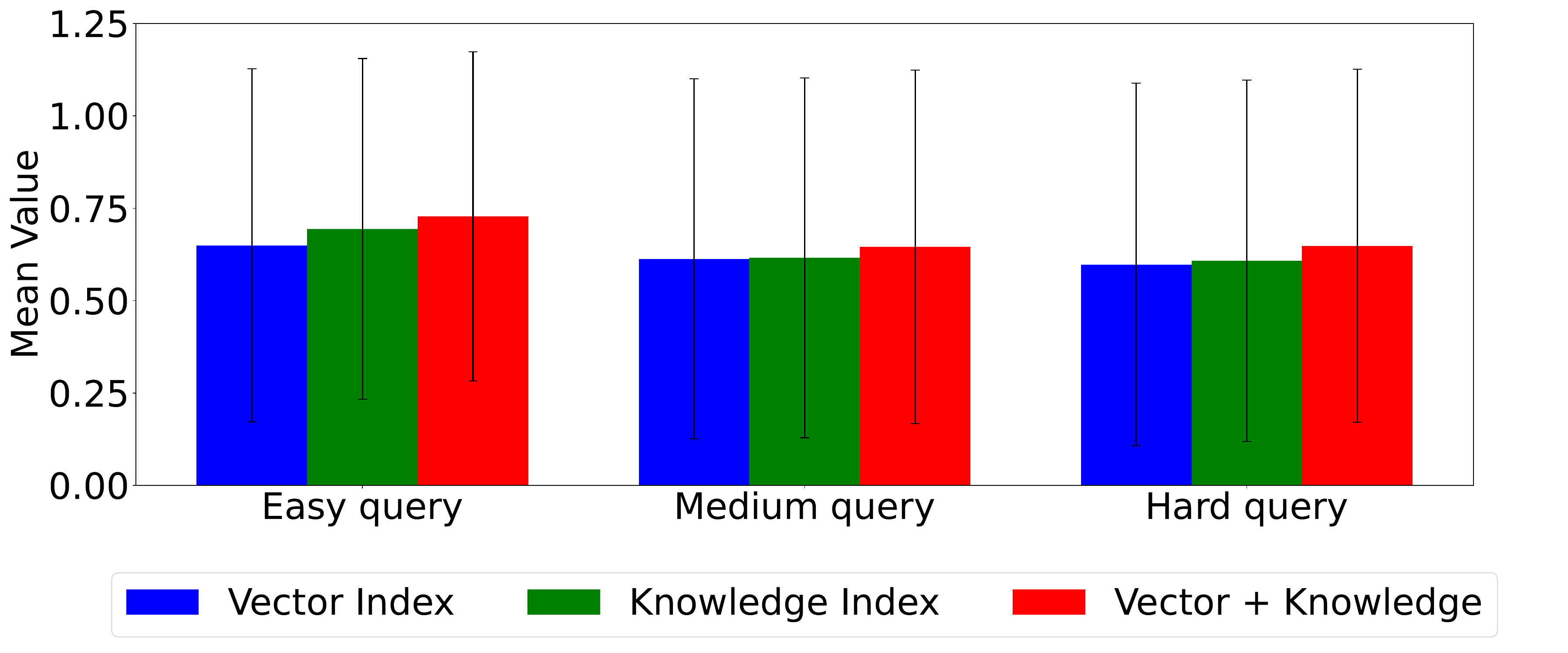}
        \label{fig:evalDLc}}
    \caption{The average quality of response based on (a) correctness, (b) relevance and (c) faithfulness for easy, medium and hard queries. Each difficulty level has 765 queries.}
    \label{fig:evalDL}
\end{figure}

The RAG based on custom index consistently outperforms vector index and KG index, especially in medium and hard queries (figure \ref{fig:evalDLb}). The Vector Index, which excels at capturing the overall semantic context, shows high relevance for easy queries where the required information is more straightforward and closely aligned with the query’s context. However, its relevance declines in medium and hard queries as the complexity of the information required increases, making it harder for the Vector Index to maintain relevance solely based on semantic similarity. The Knowledge Index, while slightly less effective for easy queries, performs more consistently across medium and hard queries, benefiting from its focus on specific entities and relationships. However, it still does not achieve the same level of relevance as the Custom Index, which integrates the strengths of both methods to maintain and even improve relevance as query complexity increases.

After aggregating across difficulty levels and keywords occurrences, the RAG based on custom index maintains the highest relevance, reflecting its ability to balance semantic context and factual precision (figure A1(b)). The Vector Index performs well for easier queries, where context plays a significant role in determining relevance, but struggles as query complexity increases. The Knowledge Index, with its steady relevance across different queries, shows that it can consistently target relevant entities, but it lacks the broader contextual understanding needed for the highest relevance in more complex scenarios. The Custom Index’s superior performance in relevance across all query sets demonstrates its effectiveness in ensuring that responses are not only accurate but also contextually aligned with the query’s intent.

\subsection{Faithfulness or Hallucination}
The RAG based on custom index consistently exhibits the lowest hallucination rates, especially in medium and hard queries (figure \ref{fig:evalDLc}). This low rate of hallucinations is a result of the combined approach’s ability to leverage both the broad semantic understanding of the Vector Index and the precise factual grounding of the Knowledge Index. The Vector Index, while improving in faithfulness for easy and medium queries, exhibits slightly higher hallucination rates for hard queries. This tendency is due to its reliance on semantic similarity, which can sometimes lead to the generation of contextually plausible but factually incorrect responses, particularly in complex or nuanced queries. The Knowledge Index, which generally has lower hallucination rates due to its structured, entity-focused retrieval approach, still faces challenges in hard queries where maintaining factual accuracy is more difficult. The Custom Index’s ability to minimize hallucinations across all levels of query difficulty underscores its robustness in providing reliable and accurate information, even when the queries are complex and challenging.

After aggregating across difficulty levels and keywords occurrences, the RAG based on custom index shows the lowest hallucination rates, indicating its effectiveness in generating responses that are both contextually appropriate and factually accurate (figure A1(c)). This is consistent with observations in figure \ref{fig:evalDLc}.

Further evaluation results were presented in figures A2-A7 to understand performances at different level of keywords occurrences. The results are generally consistent with those in figures \ref{fig:evalDL} and A1. We also did qualitative evaluation by inviting researchers of a CPG company to evaluate different components of \textbf{ScienceSage} using multimodal documents of different subjects. The feedback has been mostly positive on the quality of generated reports and short responses from the RAG.    
    
\section{Discussion}
The MVP web app was deployed on a GPU server on premise ( NVIDIA GeForce RTX 2080, 11GB memory). It has been used hundreds of times per day on average by users across business units at a consumer packaged goods company to extract insights from Voice of Science data including journal articles and technical reports etc. It has substantially sped up research projects across organizations involving quick iteration of ideas, information search and summary etc. Users actively build, store, update and query their knowledge bases using \textbf{ScienceSage} at unprecedented speed, magnitude and scope.

\textbf{ScienceSage} has been developed based on open source codes from Tavily Search and GPTResearcher. While it has similar features to commercial and open source tools like Perplexity, Tavily Search, or GPTResearcher, it distinguishes itself by using a common set a knowledge bases (i.e. vector and multimodal databases) to connect all three functions: generating research report, RAG for textual documents and RAG for multimodal data. The knowledge bases are built from multiple sources of multimodal data such as journal articles, patents and internal technical documents including text, image, audio and video. These persistent knowledge bases are continuously updated by users to answer their questions either with a short answer through a RAG or generating a long research report with references. \textbf{ScienceSage} is not designed to offer substantial algorithmic innovation for generative AI. It aims to assemble the best and the most efficient GenAI packages to significantly boost efficiency of our R\&D activities and also to drive product innovation through diverse and interdisciplinary knowledge bases.           

One challenge is to manage different embedding based databases such as Weaviate, Nebula Graph and LanceDB. We selected these databases based on exploratory learning. Ideally, we just need one database to hold vector index, KG index and multimodal index. To our best knowledge, we don't have an open source embedding based database to efficiently store and query all these indices yet.

The compatibility of different version of LLM packages is another challenge affecting across platform deployment. We understand platforms like LangChain and LlamaIndex are quickly evolving, but hope each platform can consider more backward compatibility and compatibility with each other.

Our improvement plan includes scaling up \textbf{ScienceSage} in cloud environment, capturing tacit knowledge from domain experts and adding additional data sources such as patent databases and additional databases of journal articles. We also plan to test and add functionalities to conduct graph analytics and graph machine learning on the extracted knowledge graphs, and to generate a podcast from a knowledge base.

We are also working to improve and evaluate the quality of generated report using methods proposed in \textbf{STORM} and \textbf{CO-STORM} \citep{shao2024assisting, jiang2024unknownunknowns}. Other than generating Wiki style articles and generic research report, we intend to adapt \textbf{STORM} to generate knowledge base grounded long documents with specific formats such as product feasibility and quality reports. Further research and consideration includes using framework like \textbf{SciSafeEval} to evaluate the safety and risk for potential misuse of AI in science \citep{li2024scisafeeval, he2023control}.

\section{Conclusion}
\textbf{ScienceSage} is a distinctive GenAI application to empower users to build, store, update and query knowledge bases at disruptive speed, magnitude and scope. Users can build the KB using their own multimodal data or searching the internet for latest information. It provides either a structural report or a short answer to user's research question. RAG of \textbf{ScienceSage} supports vector index, KG index, custom index and multimodal index. The RAG based on the custom index generally offers superior performance at small compromise of speed.  

\section{Acknowledgments}
We thank our colleagues Ming Chen and George Gabone for their technical help. We acknowledge the Turing Project team at P\&G for enabling access to ChatPG/ChatGPT API in Azure. 

\clearpage
\bibliography{aaai25}

\clearpage
\setcounter{secnumdepth}{1} 
\appendix
\counterwithin{figure}{section}
\counterwithin*{equation}{section}
\counterwithin*{figure}{section}
\renewcommand\theequation{\thesection\arabic{equation}}
\renewcommand\thefigure{\thesection\arabic{figure}}
\addcontentsline{toc}{section}{Appendices}
\section{Appendix: Build Your Knowledge Base Using GenAI Powered ScienceSage} \label{appendix}

\begin{figure*}[!ht] 
    \centering
    \subfigure[Correctness of response]{
        \includegraphics[width=0.3\linewidth]{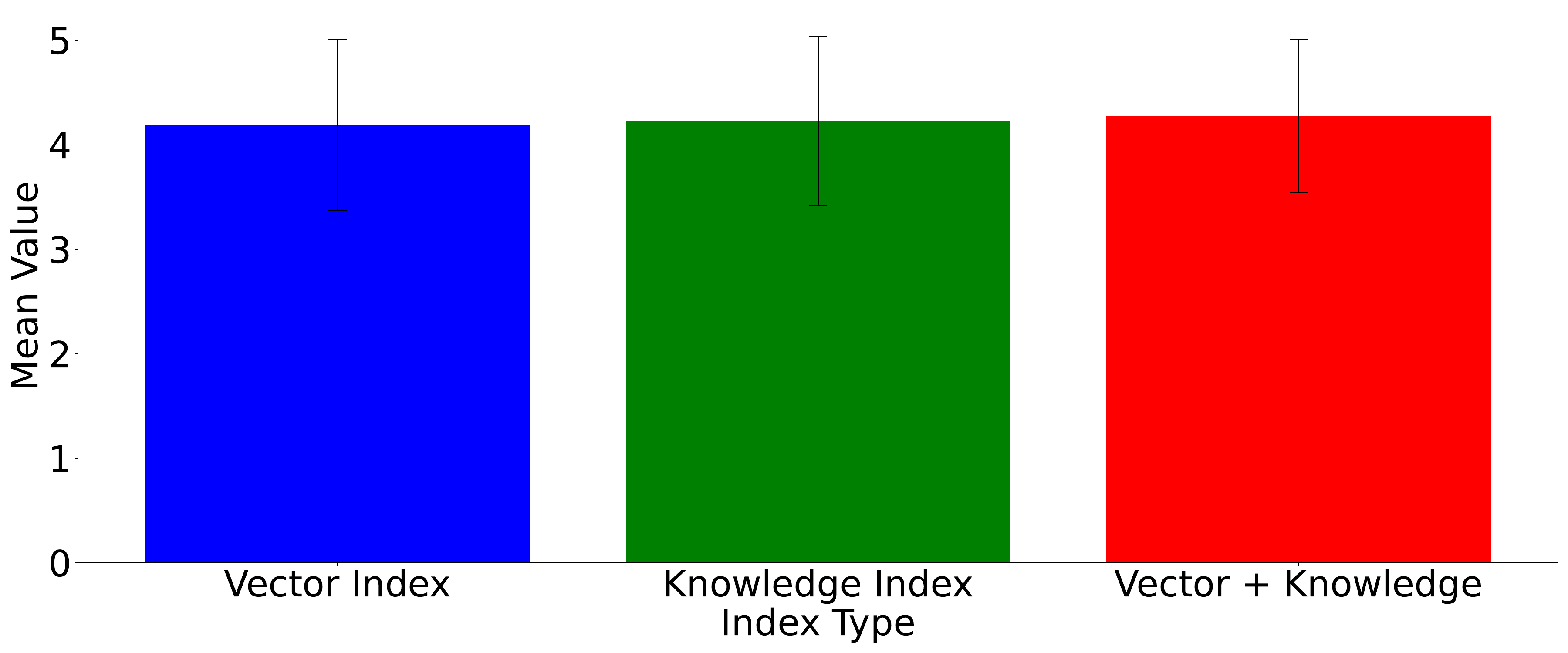}
        \label{fig:evalDLagga}}
    \subfigure[Relevance of response]{
        \includegraphics[width=0.3\linewidth]{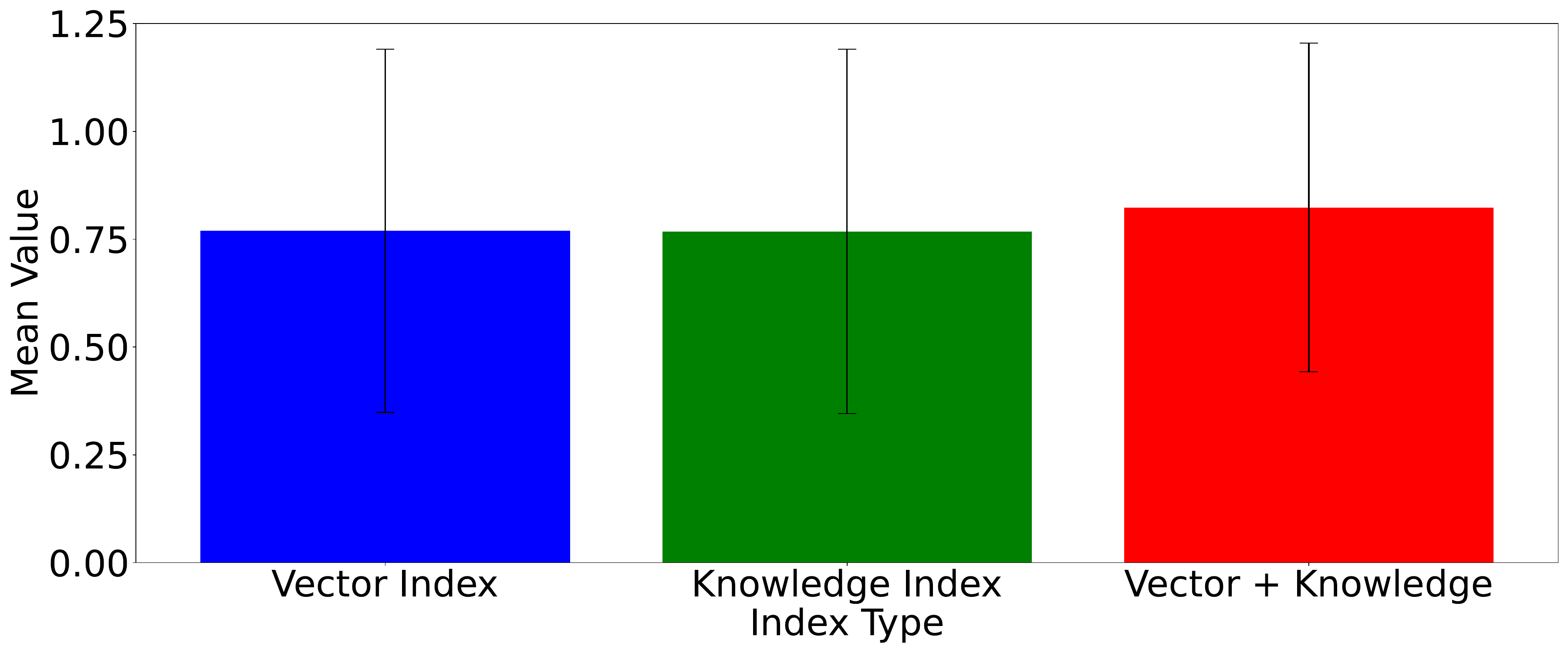}
        \label{fig:evalDLaggb}}
    \subfigure[Faithfulness of response]{
        \includegraphics[width=0.3\linewidth]{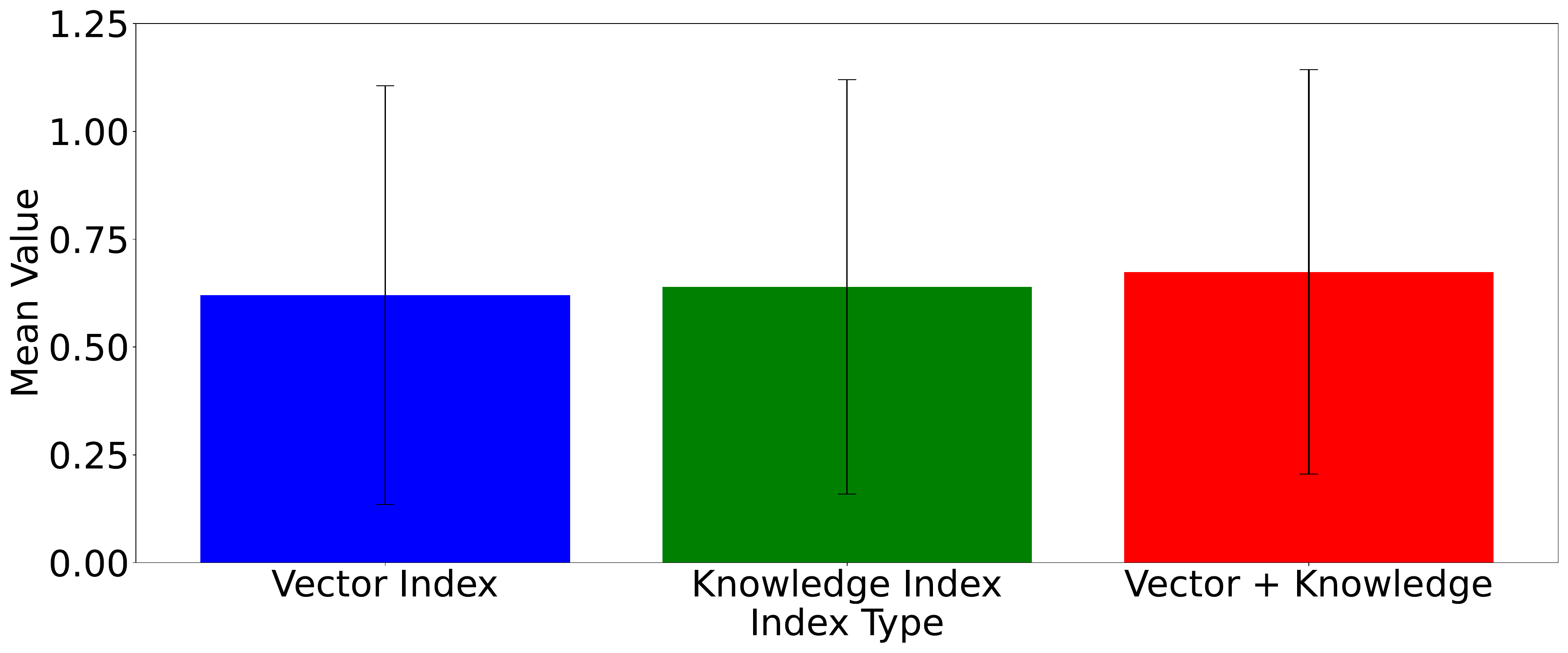}
        \label{fig:evalDLaggc}}
    \caption{The average quality of response based on (a) correctness, (b) relevance and (c) faithfulness for all queries. There are 2295 queries.}
    \label{fig:evalDLagg}
\end{figure*}

\begin{figure*}[!ht] 
    \centering
    \subfigure[Correctness of response]{
        \includegraphics[width=0.6\linewidth]{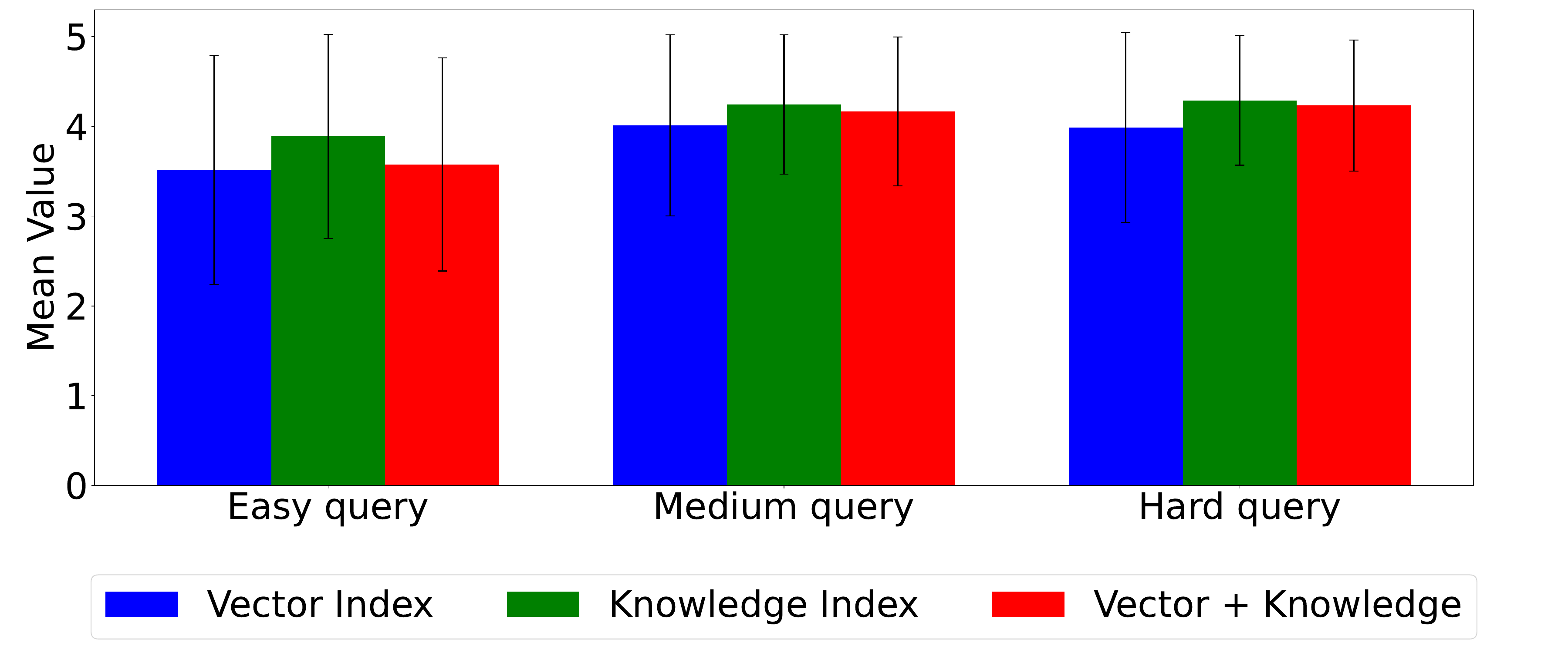}
        \label{fig:aveComHCosine}}
    \subfigure[Relevance of response]{
        \includegraphics[width=0.6\linewidth]{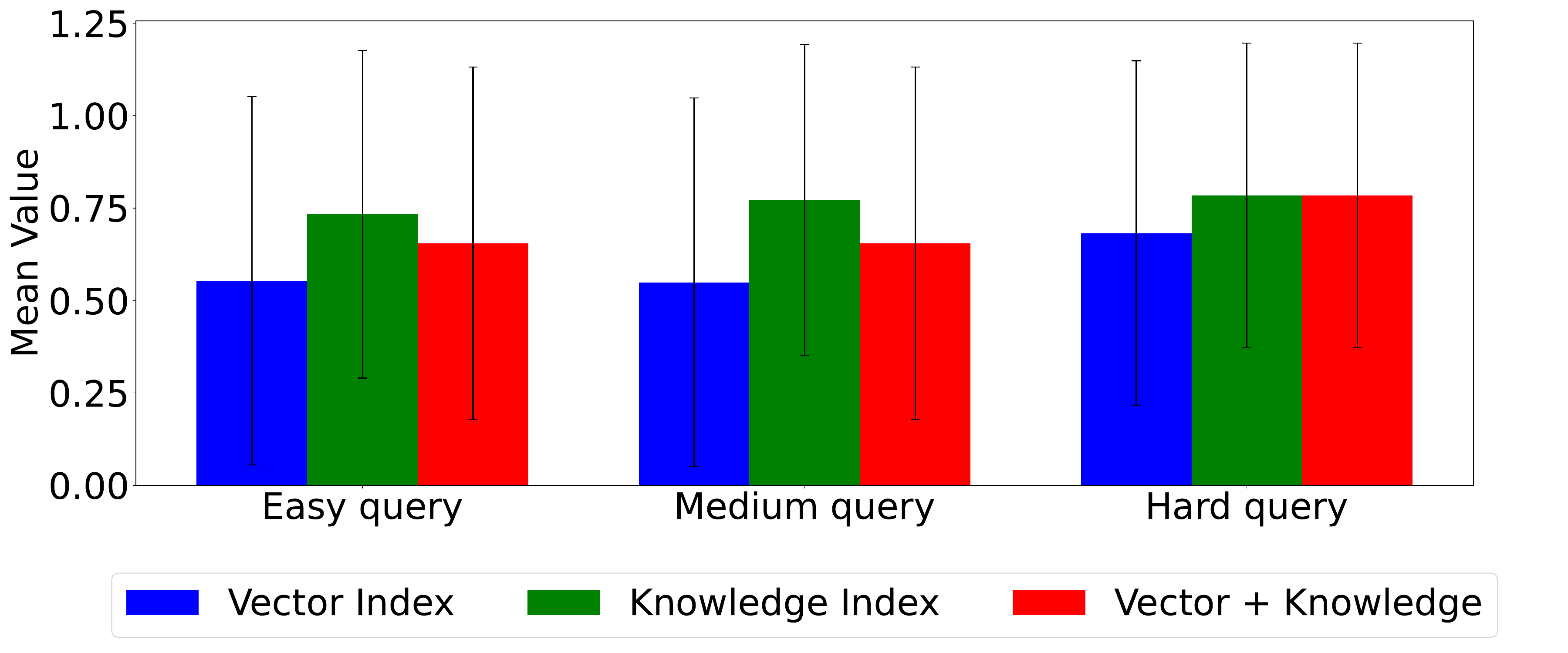}
        \label{fig:aveComHEuclidean}}
    \subfigure[Faithfulness of response]{
        \includegraphics[width=0.6\linewidth]{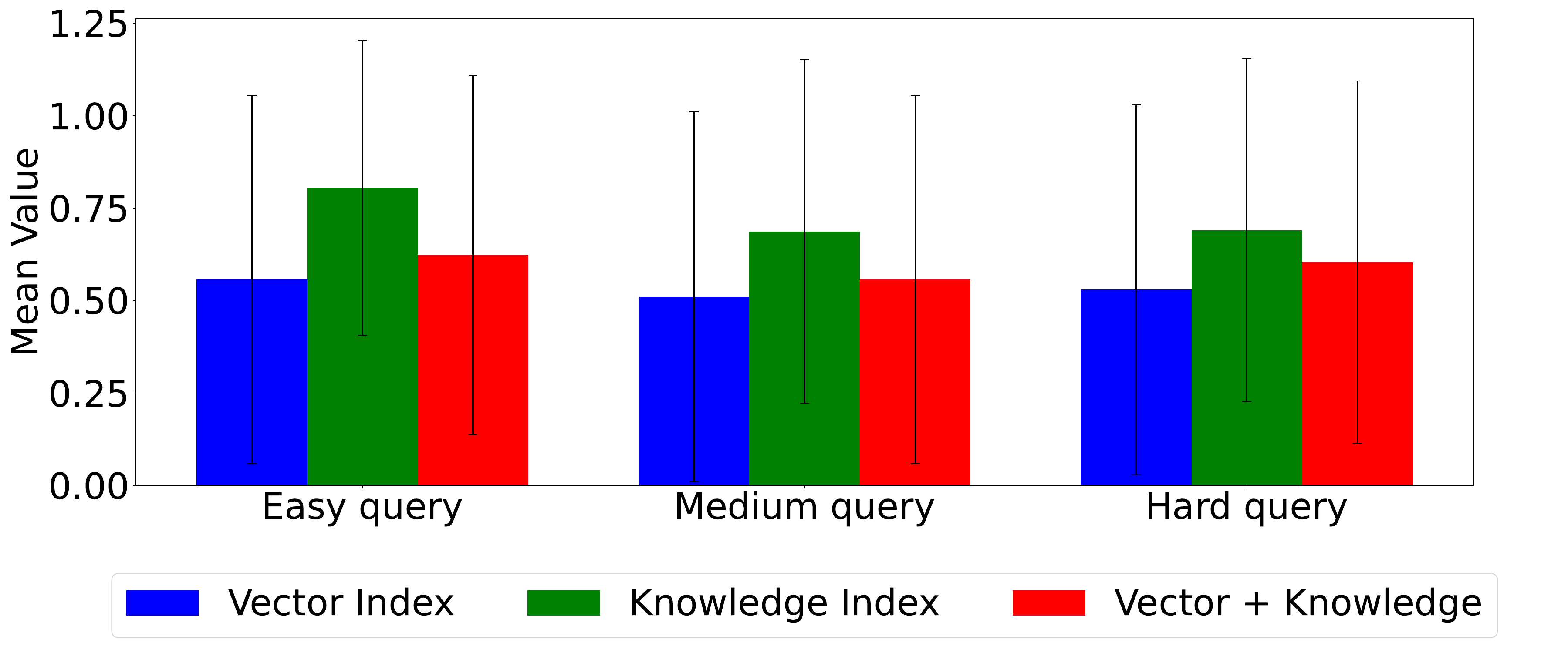}
        \label{fig:aveComHCanberra}}
    \caption{The average quality of response based on (a) correctness, (b) relevance and (c) faithfulness for easy, medium and hard queries with less keywords in the queries. Each difficulty level has 255 queries.}
    \label{fig:evalDLLowKW}
\end{figure*}

\begin{figure*}[!ht] 
    \centering
    \subfigure[Correctness of response]{
        \includegraphics[width=0.3\linewidth]{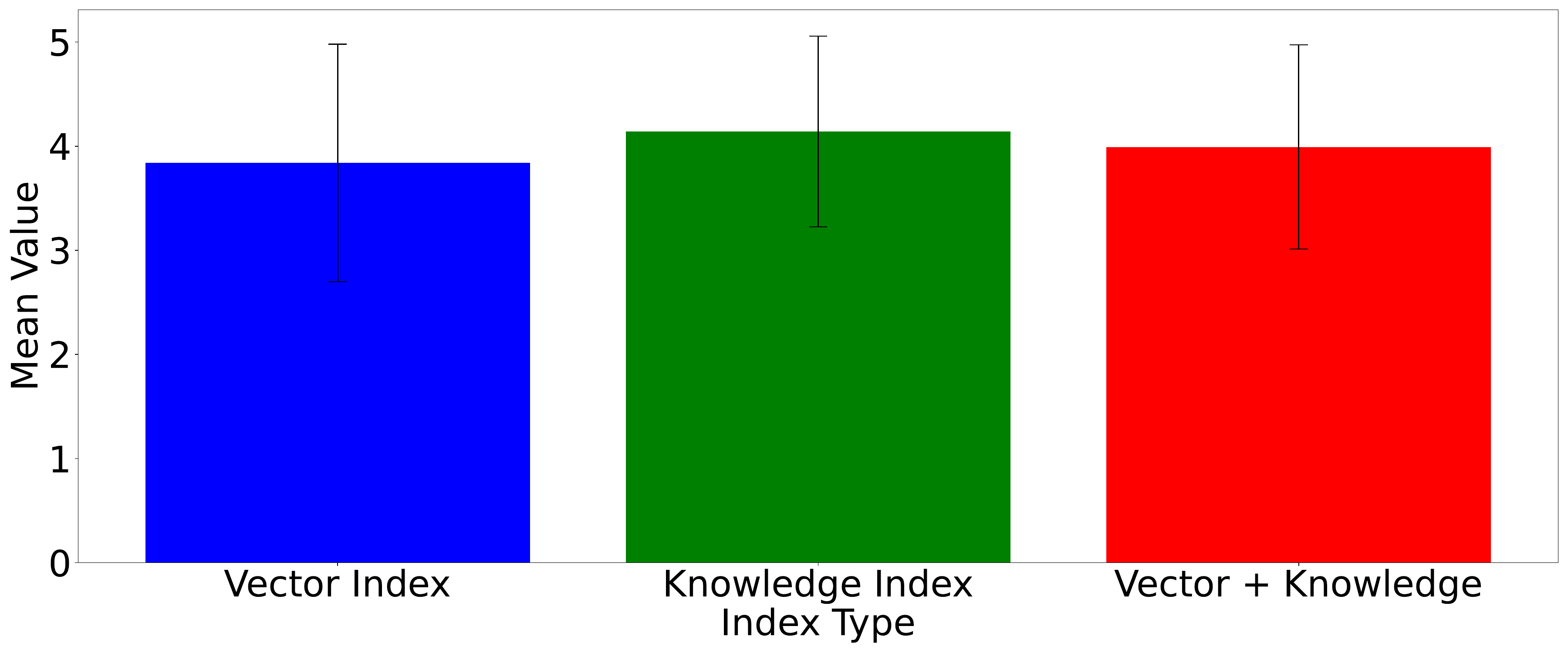}
        \label{fig:aveComHCosine}}
    \subfigure[Relevance of response]{
        \includegraphics[width=0.3\linewidth]{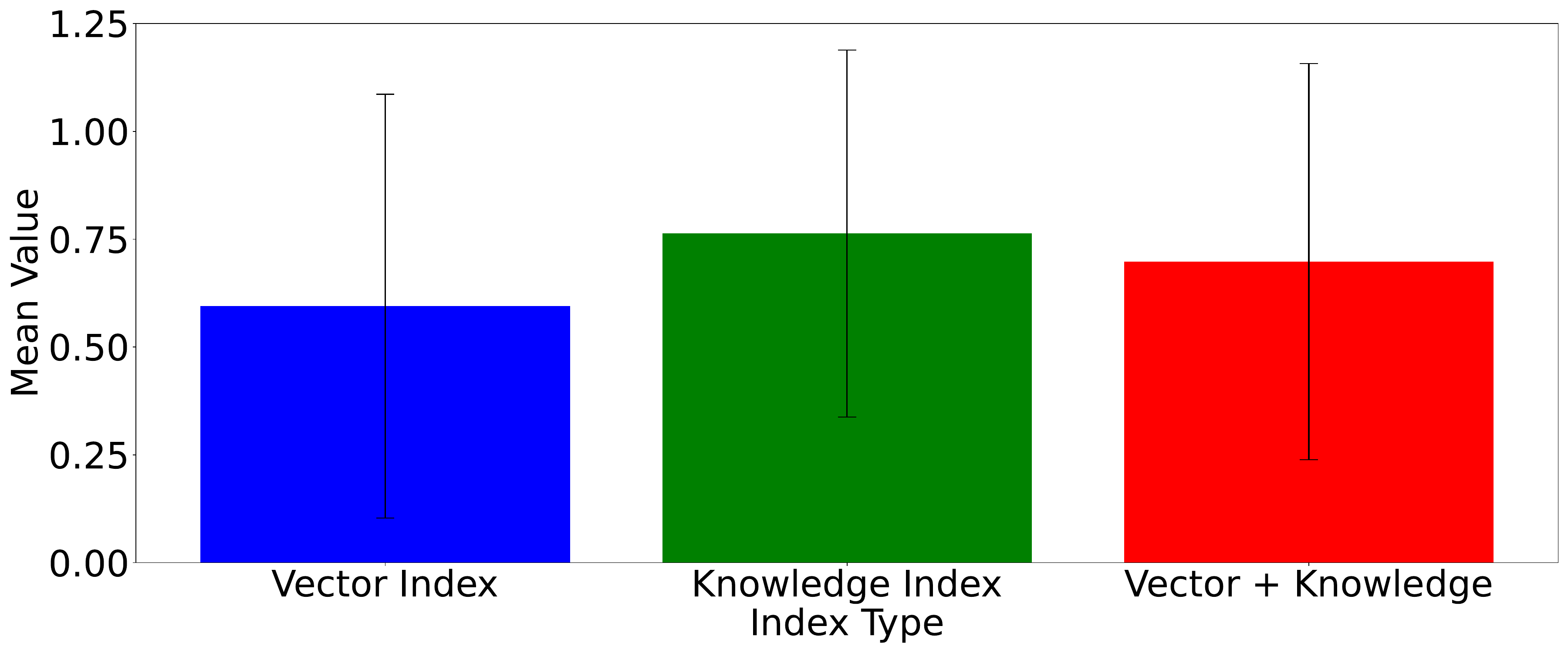}
        \label{fig:aveComHEuclidean}}
    \subfigure[Faithfulness of response]{
        \includegraphics[width=0.3\linewidth]{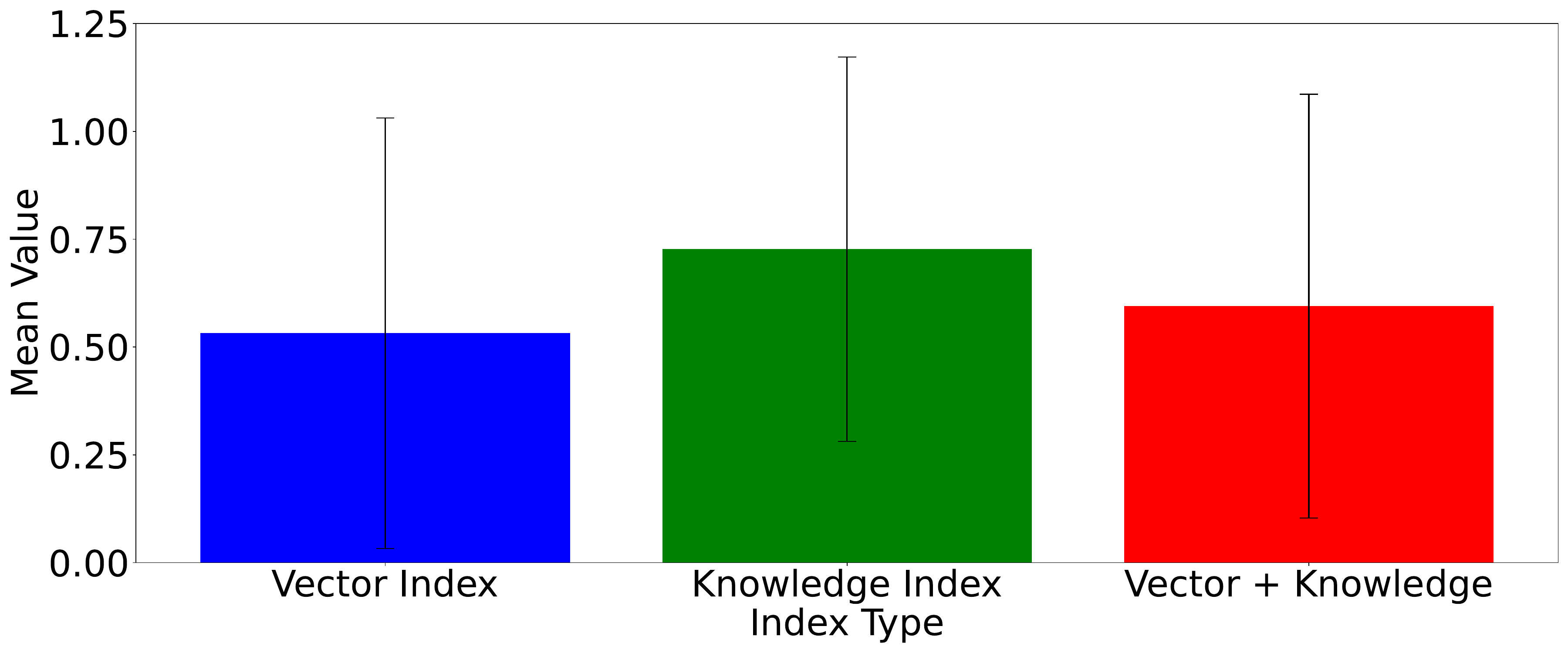}
        \label{fig:aveComHCanberra}}
    \caption{The average quality of response based on (a) correctness, (b) relevance and (c) faithfulness for all queries with less keywords in the queries. There are 765 queries.}
    \label{fig:evalDLLowKWagg}
\end{figure*}

\begin{figure*}[!ht] 
    \centering
    \subfigure[Correctness of response]{
        \includegraphics[width=0.6\linewidth]{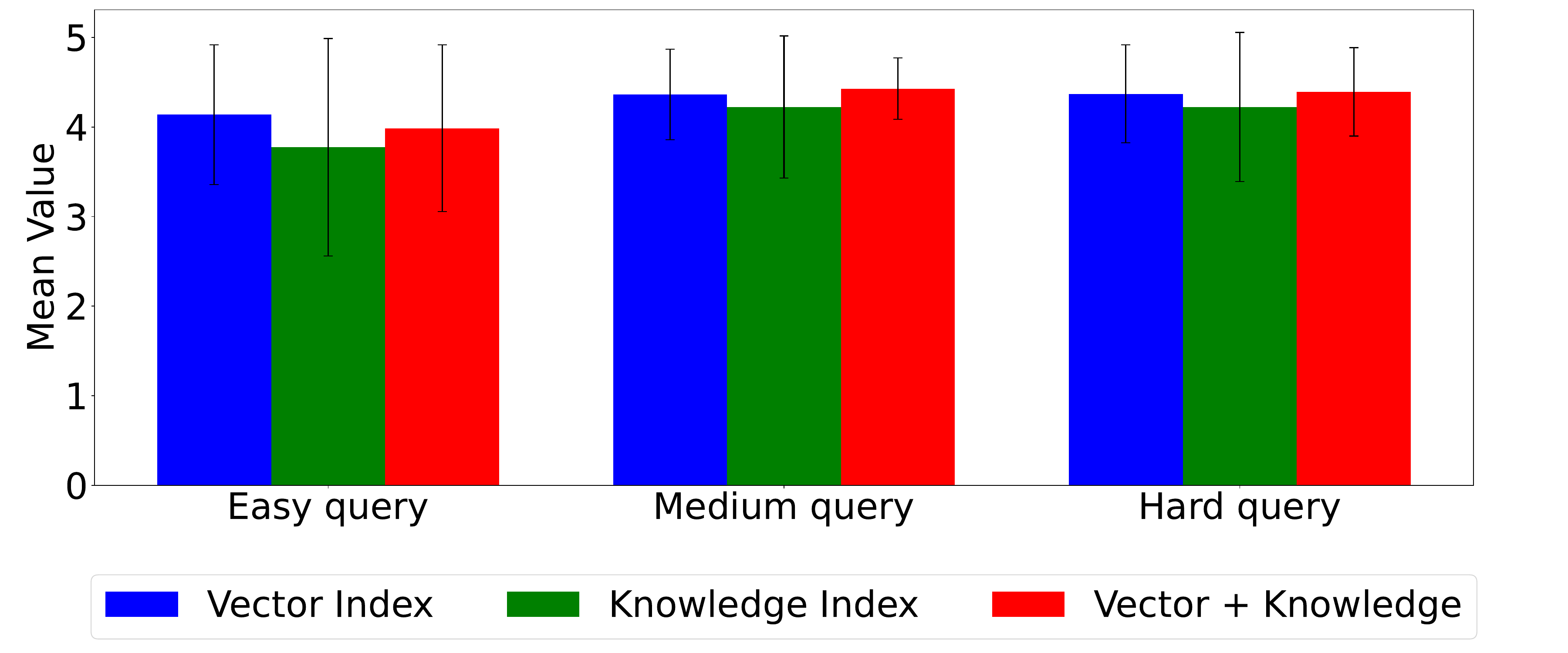}
        \label{fig:aveComHCosine}}
    \subfigure[Relevance of response]{
        \includegraphics[width=0.6\linewidth]{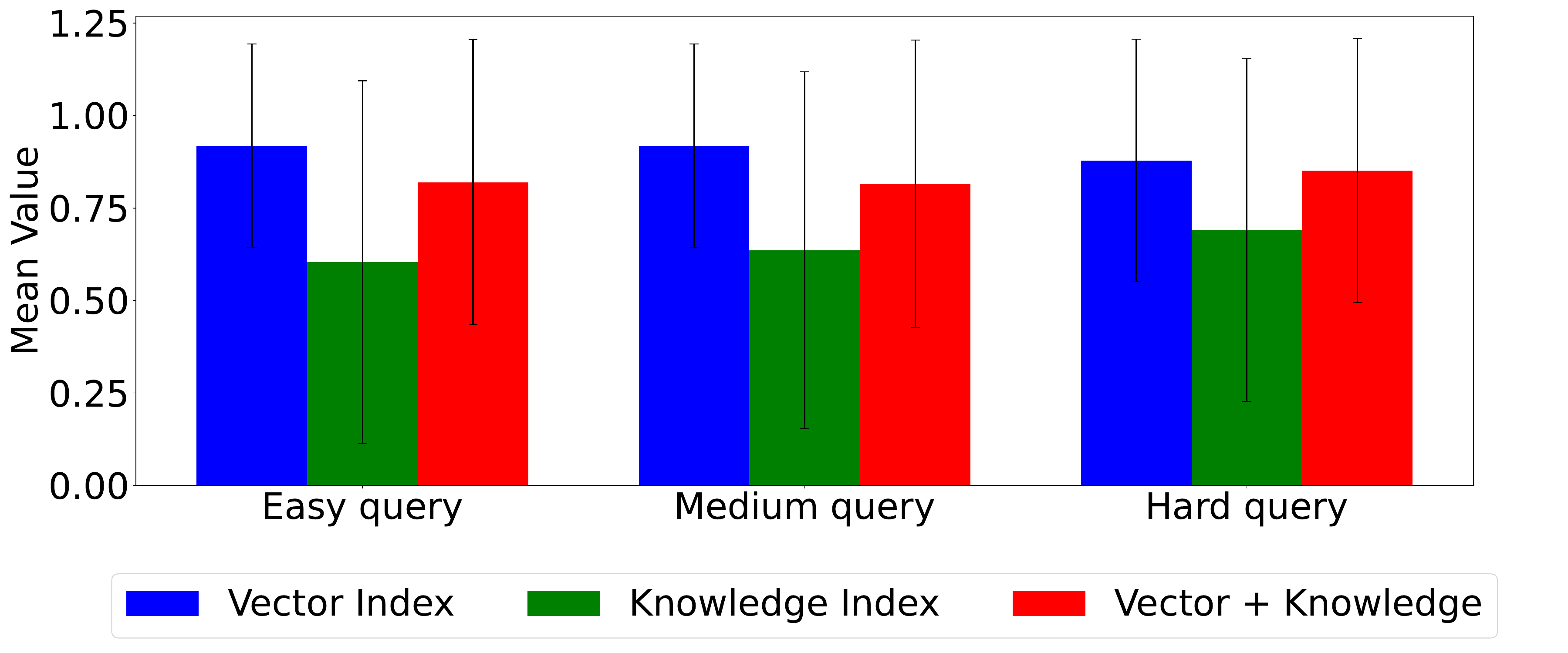}
        \label{fig:aveComHEuclidean}}
    \subfigure[Faithfulness of response]{
        \includegraphics[width=0.6\linewidth]{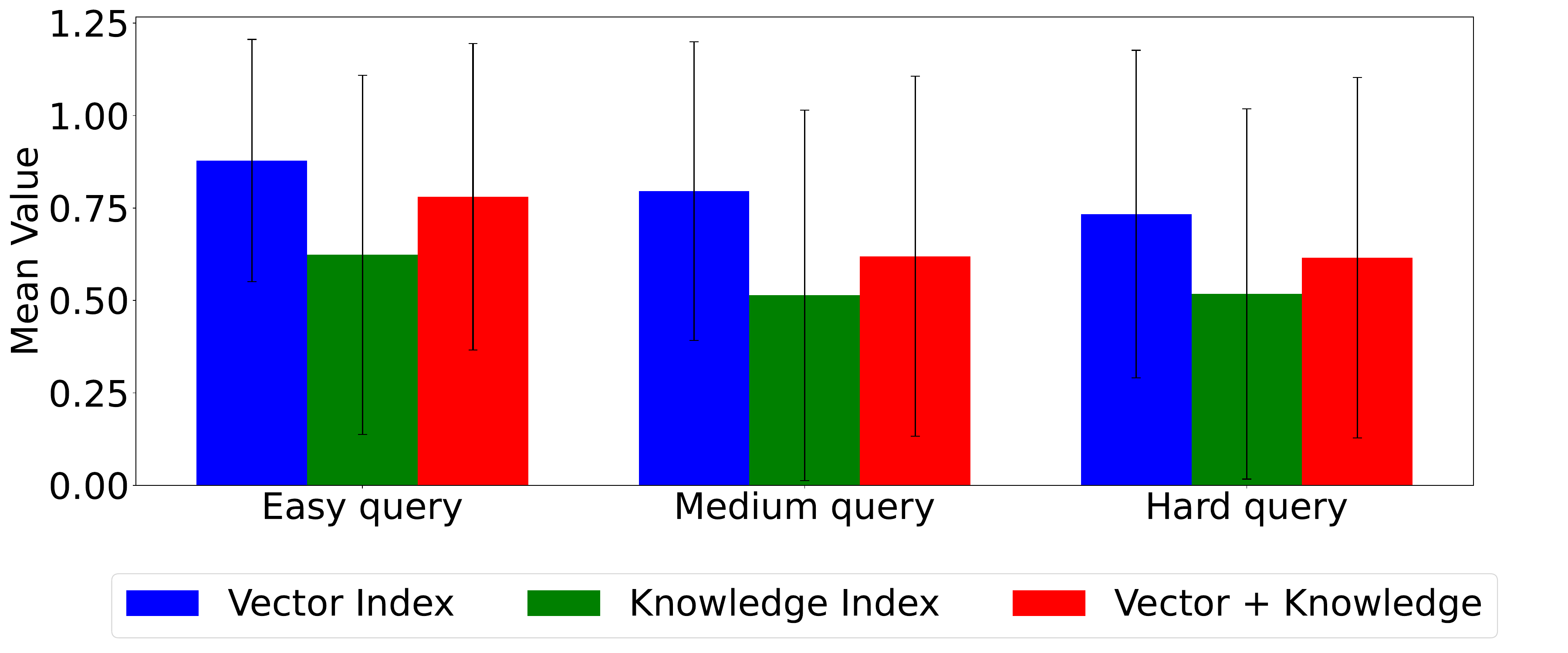}
        \label{fig:aveComHCanberra}}
    \caption{The average quality of response based on (a) correctness, (b) relevance and (c) faithfulness for easy, medium and hard queries with a norm number of keywords in the queries. Each difficulty level has 255 queries.}
    \label{fig:evalDLMedKW}
\end{figure*}

\begin{figure*}[!ht] 
    \centering
    \subfigure[Correctness  of response]{
        \includegraphics[width=0.3\linewidth]{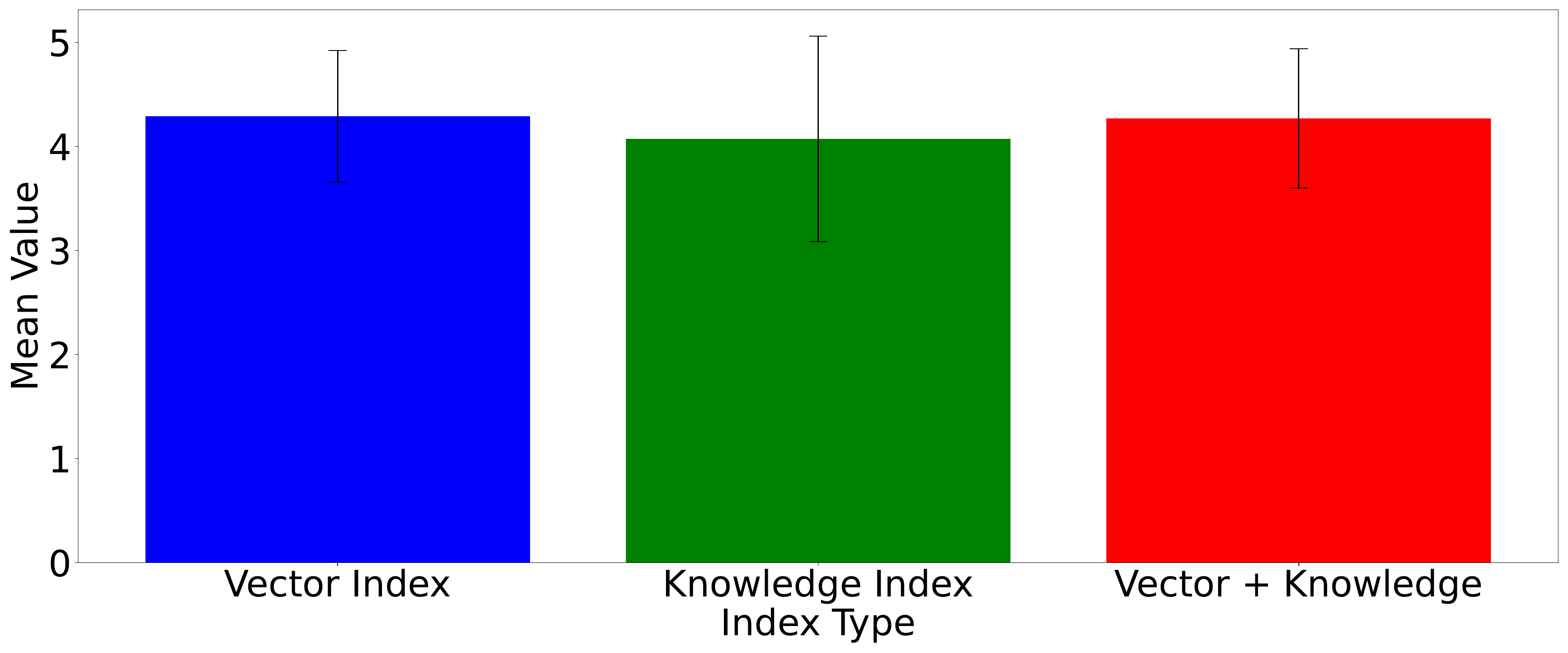}
        \label{fig:aveComHCosine}}
    \subfigure[Relevance  of response]{
        \includegraphics[width=0.3\linewidth]{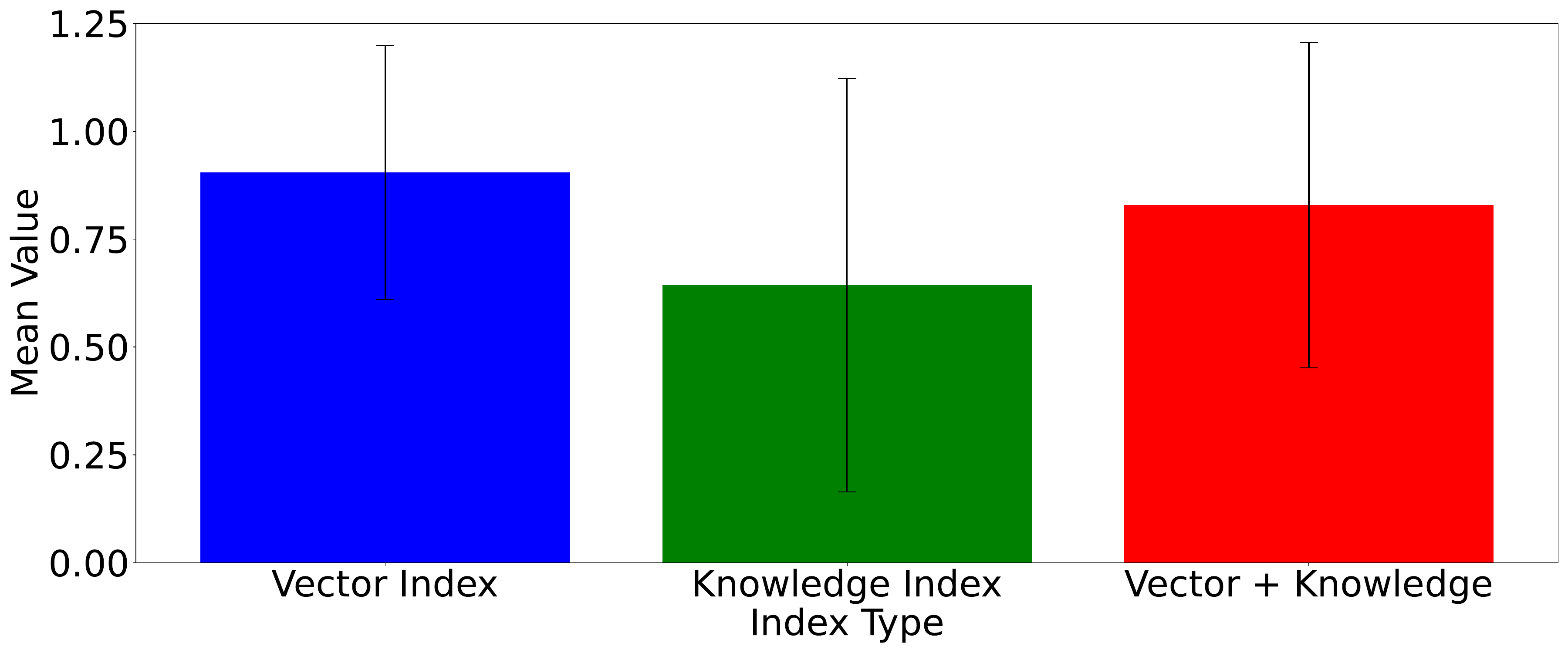}
        \label{fig:aveComHEuclidean}}
    \subfigure[Faithfulness  of response]{
        \includegraphics[width=0.3\linewidth]{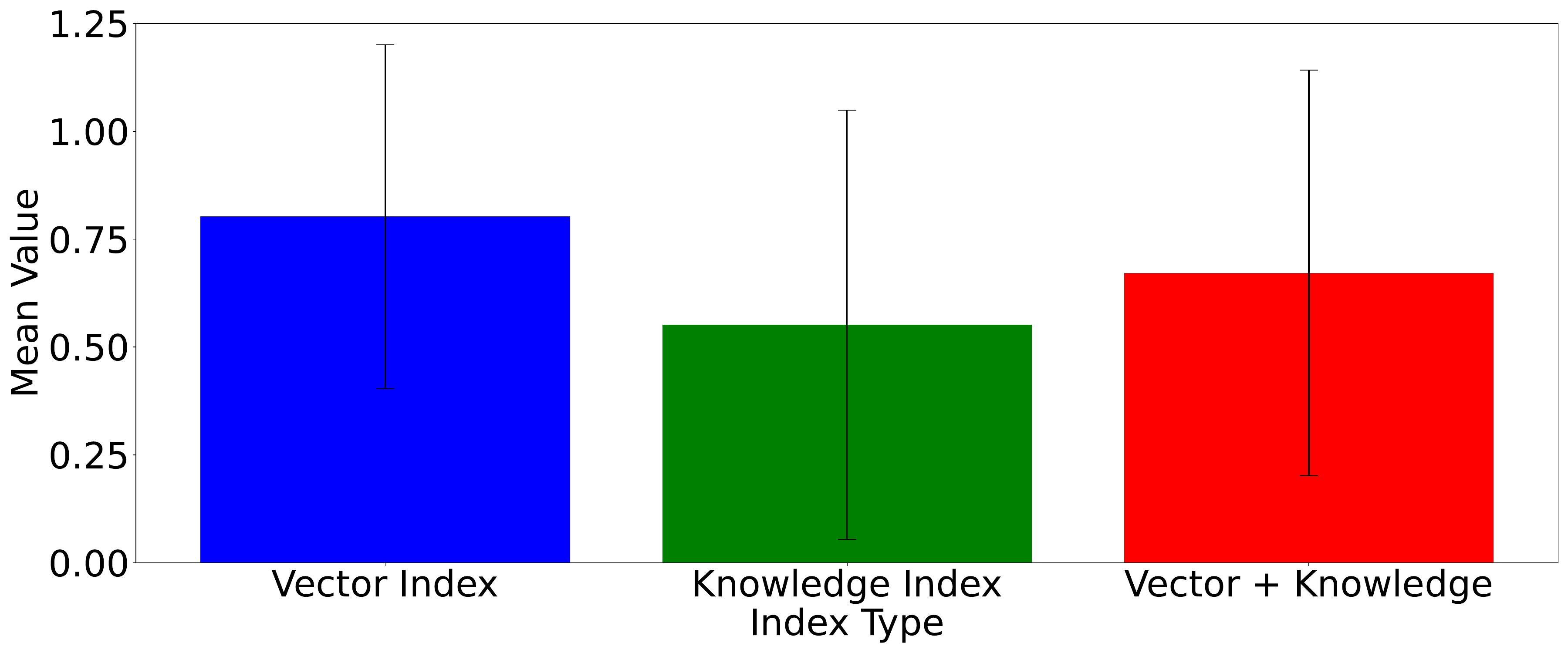}
        \label{fig:aveComHCanberra}}
    \caption{The average quality of response based on (a) correctness, (b) relevance and (c) faithfulness for all queries with a norm number of keywords in the queries. There are 765 queries.}
    \label{fig:evalDLMedKWagg}
\end{figure*}

\begin{figure*}[!ht] 
    \centering
    \subfigure[Correctness of response]{
        \includegraphics[width=0.6\linewidth]{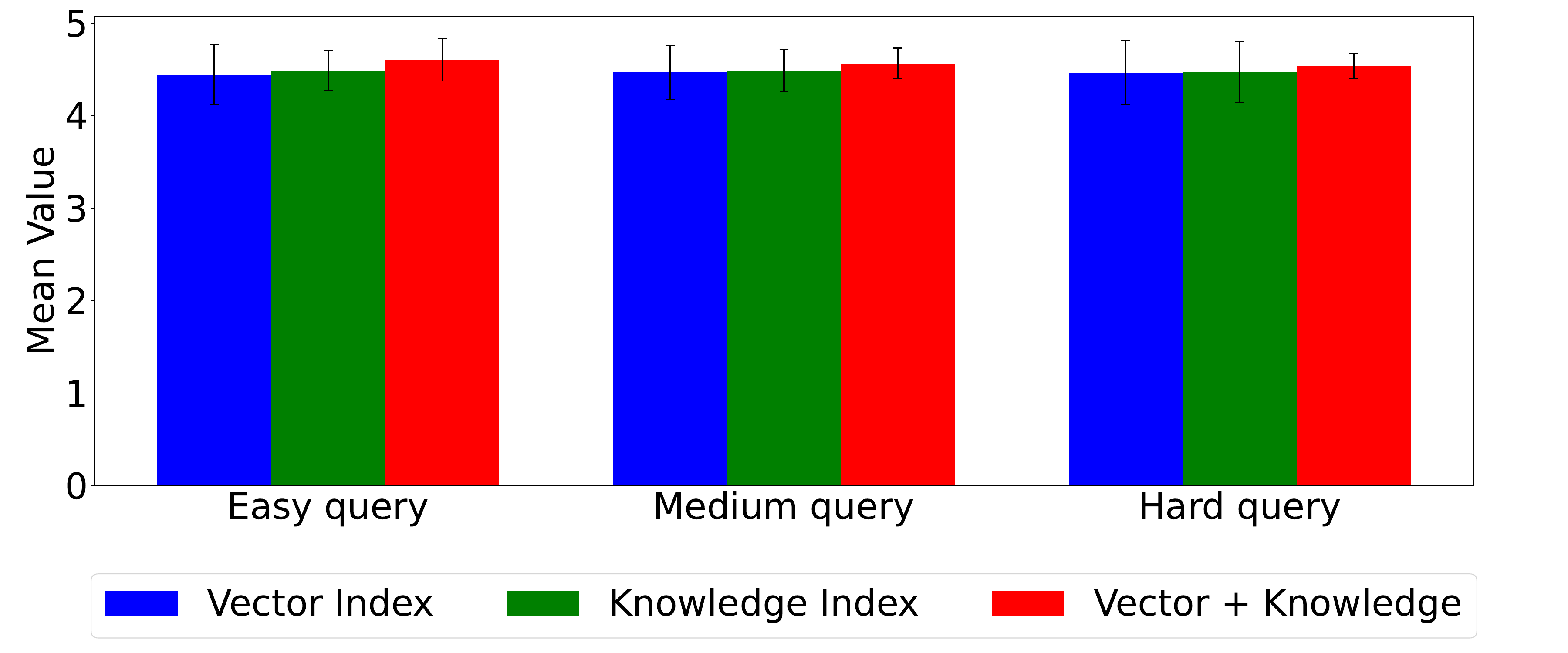}
        \label{fig:aveComHCosine}}
    \subfigure[Relevance of response]{
        \includegraphics[width=0.6\linewidth]{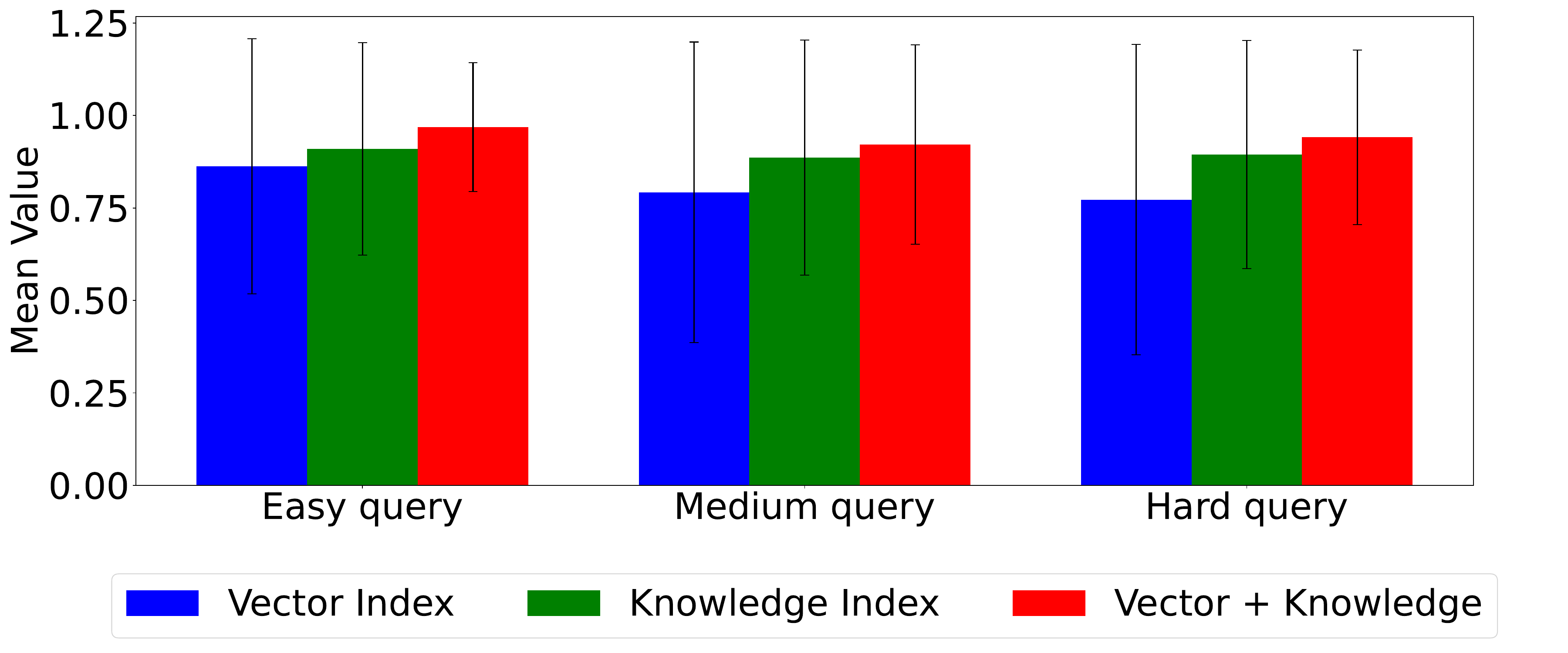}
        \label{fig:aveComHEuclidean}}
    \subfigure[Faithfulness of response]{
        \includegraphics[width=0.6\linewidth]{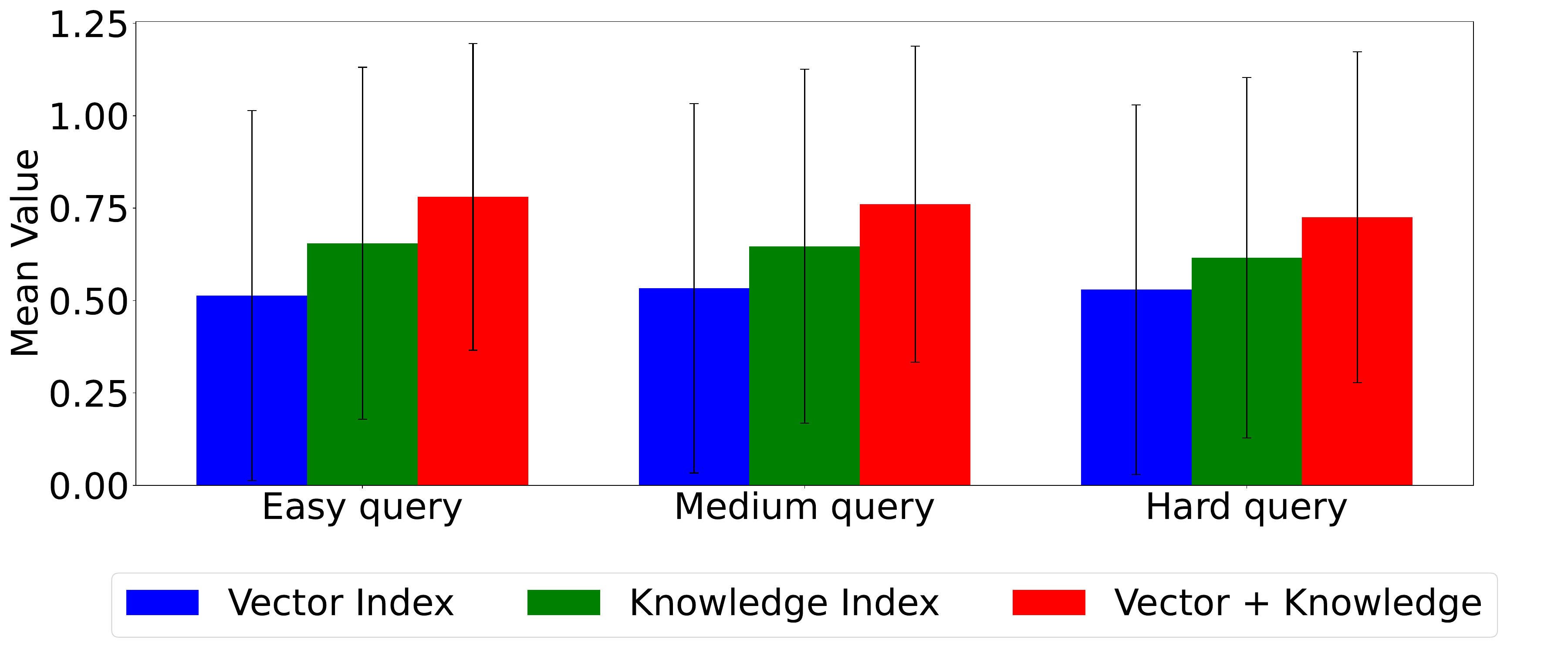}
        \label{fig:aveComHCanberra}}
    \caption{The average quality of response based on (a) correctness, (b) relevance and (c) faithfulness for easy, medium and hard queries with more keywords in the queries. Each difficulty level has 255 queries.}
    \label{fig:evalDLhighKW}
\end{figure*}

\begin{figure*}[!ht] 
    \centering
    \subfigure[Correctness]{
        \includegraphics[width=0.3\linewidth]{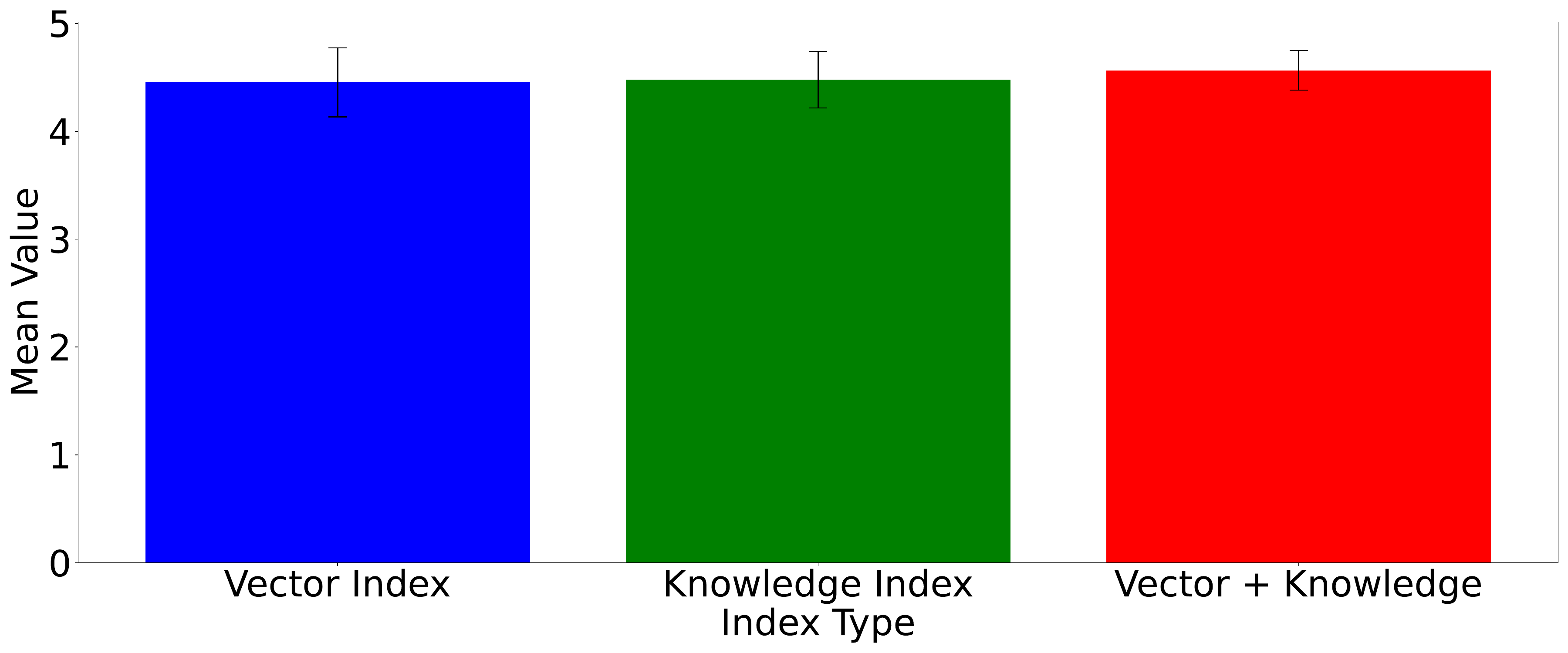}
        \label{fig:aveComHCosine}}
    \subfigure[Relevance of response]{
        \includegraphics[width=0.3\linewidth]{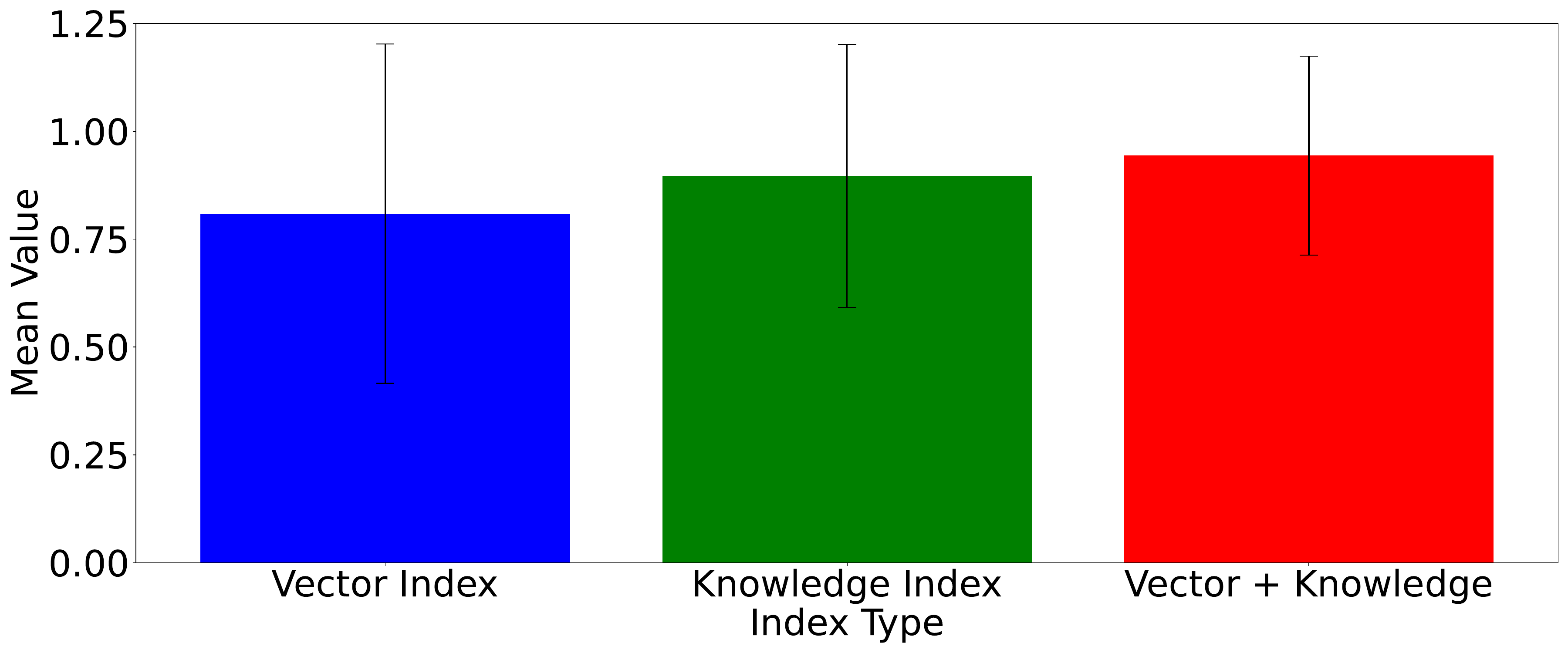}
        \label{fig:aveComHEuclidean}}
    \subfigure[Faithfulness]{
        \includegraphics[width=0.3\linewidth]{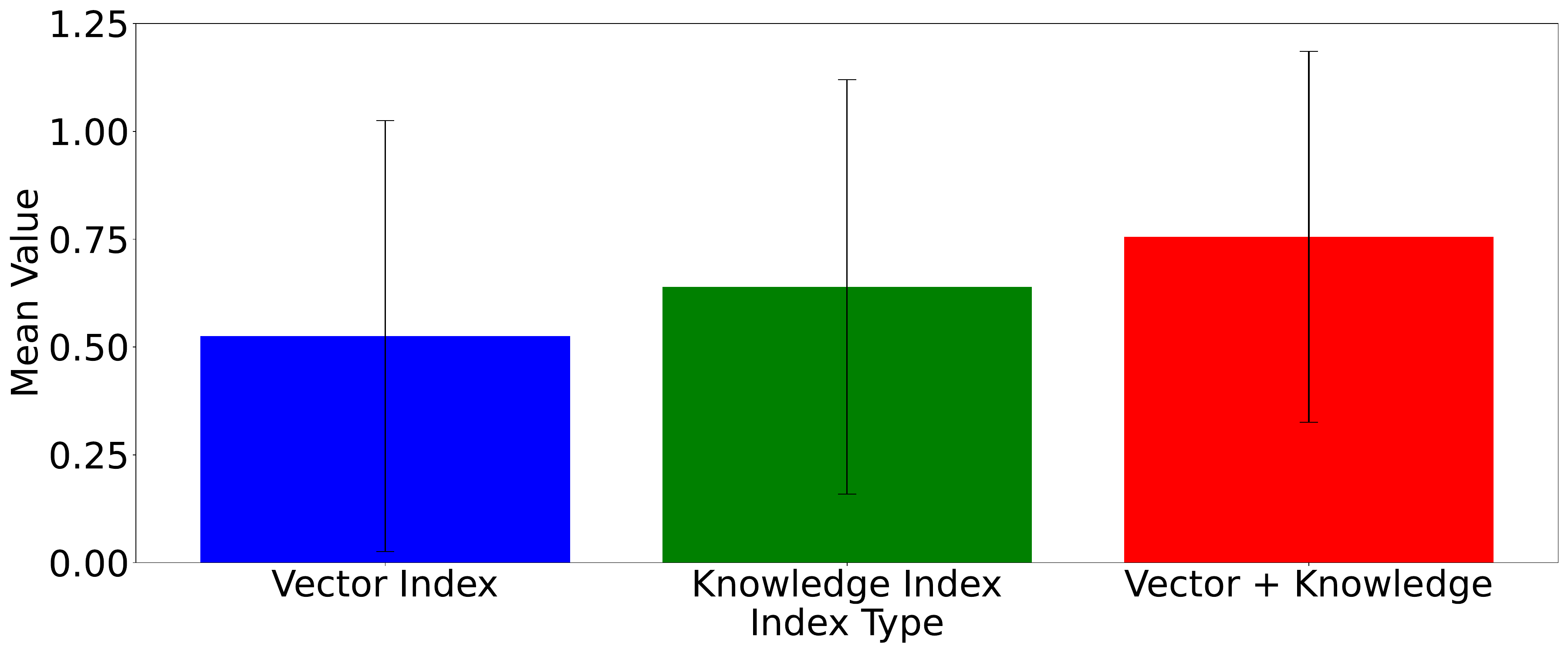}
        \label{fig:aveComHCanberra}}
    \caption{The average quality of response based on (a) correctness, (b) relevance and (c) faithfulness for all queries with more keywords in the queries. There are 765 queries.}
    \label{fig:evalDLhighKWagg}
\end{figure*}

\end{document}